\documentclass[12pt]{iopart}

\usepackage{iopams}
\usepackage{setstack}
\usepackage{graphicx}
\usepackage[caption = false]{subfig}
\usepackage{color}
\usepackage[numbers,sort&compress]{natbib}

\newcommand{\para}{{\mkern4mu\vphantom{\perp}\vrule depth 0pt\mkern3mu\vrule depth 0pt\mkern5mu}}

\begin{document}

\title[]{Asymmetric pattern formation in microswimmer suspensions induced by orienting fields}

\author{Henning Reinken$^1$, Sebastian Heidenreich$^2$, Markus B\"ar$^2$, Sabine H.L. Klapp$^1$}

\address{$^1$Institute for Theoretical Physics, Technische Universit\"at Berlin,
Hardenbergstr. 36, D-10623, Berlin, Germany}
\address{$^2$Department of Mathematical Modelling and Data Analysis,
Physikalisch-Technische Bundesanstalt Braunschweig und Berlin, Abbestr. 2-12,
D-10587 Berlin, Germany}
\ead{henning.reinken@itp.tu-berlin.de}

\begin{abstract}
This paper studies the influence of orienting external fields on pattern formation, particularly mesoscale turbulence, in microswimmer suspensions.
To this end, we apply a hydrodynamic theory that can be derived from a microscopic microswimmer model~[H.~Reinken \textit{et~al.}, Phys. Rev. E \textbf{97}, 022613 (2018)].
The theory combines a dynamic equation for the polar order parameter with a modified Stokes equation for the solvent flow.
Here, we extend the model by including an external field that exerts an aligning torque on the swimmers (mimicking the situation in chemo-, photo-, magneto- or gravitaxis). 
Compared to the field-free case, the external field breaks the rotational symmetry of the vortex dynamics and leads instead to strongly asymmetric, traveling stripe patterns, as demonstrated by numerical solution and linear stability analysis.
We further analyze the emerging structures using a reduced model which involves only an (effective) microswimmer velocity field.
This model is significantly easier to handle analytically, but still preserves the main features of the asymmetric pattern formation.
We observe an underlying transition between a square vortex lattice and a traveling stripe pattern.
These structures can be well described in the framework of weakly nonlinear analysis, provided the strength of nonlinear advection is sufficiently weak.
\end{abstract}
\noindent{\it Keywords\/}: active matter, microswimmers, pattern formation, mesoscale turbulence, weakly nonlinear analysis

\submitto{\NJP}

\maketitle

\section{Introduction}

Active matter exhibits a variety of large-scale self-organized structures that arise due to the interactions between the moving constituents.
As shown in experiments, these structures are ranging from dynamical clustering~\cite{buttinoni2013dynamical,peruani2012collective} and giant number fluctuations~\cite{zhang2010collective} to vortices and swirling~\cite{schaller2010polar,sumino2012large,rabani2013collective}.
To describe and understand the fascinating collective behavior from a theoretical point of view, models on different levels of detail have been applied, from studies simulating large numbers of individual particles~\cite{vicsek1995novel,stenhammer2017role} to phenomenological approaches~\cite{simha2002hydrodynamic,ramaswamy2003active,toner2005hydrodynamics,cates2008shearing,heidenreich2011nonlinear,giomi2015geometry} standing on the opposite side of the spectrum.
To bridge the gap, many efforts have been made to derive coarse-grained hydrodynamic theories from microscopic models~\cite{saintillan2008instabilities,saintillan2009active,baskaran2008hydrodynamics,heidenreich2016hydrodynamic,reinken2018derivation}.
For a selection of recent reviews on the full scope of active matter see~\cite{lauga2009hydrodynamics,ramaswamy2010mechanics,romanczuk2012active,Marchetti2013,Elgeti2015,menzel2015tuned,zottl2016emergent,Bechinger2016,klapp2016collective}.

While many of the self-organized structures in systems of active constituents are well understood, the impact of external fields on the spatio-temporal pattern formation is mostly unexplored.
For example, the motion of biological microswimmers such as bacteria or algae cells, is strongly determined by the response to external stimuli.
These stimuli are of various origins: 
For example, swimmers react to concentration gradients (chemotaxis)~\cite{eisenbach2004chemotaxis,taktikos2011modeling}, a light source (phototaxis)~\cite{garcia2013light,martin2016photofocusing} or magnetic and gravitational fields (magnetotaxis~\cite{bazylinski2004magnetosome,waisbord2016environment,nadkarni2013comparison,popp2014polarity} and gravitaxis~\cite{fukui1985negative,tenhagen2014gravitaxis}).
The interplay between externally applied fields and internally generated motion is not only interesting from a fundamental perspective, but also essential for a variety of potential applications.
For example, external fields offer the possibility to control the suspension in order to exploit the coherent motion of the swimmers.
This includes tasks like cargo-delivery~\cite{trivedi2015bacterial,sokolov2015individual,schwarz2017hybrid} (e.g., drug transport for medical purposes~\cite{martel2009flagellated,felfoul2016magneto}), powering microfluidic devices~\cite{kaiser2014transport,sokolov2010swimming} or swimmer-induced mixing on the small scales, where high Reynolds numbers are not accessible~\cite{jalali2015microswimmer}.

In the present paper, we explore the impact of an external field on a prominent example of active pattern formation, labeled and known as mesoscale turbulence~\cite{wensink2012meso}.
This state can be observed in a variety of systems, including bacterial suspensions~\cite{dombrowski2004self,zhang2009swarming,aranson2012physical,lushi14fluid} and Janus particles~\cite{nishiguchi2015mesoscopic}. 
Mesoscale turbulence is characterized by chaotic vortex structures similar to inertial turbulence occurring in passive fluids at high Reynolds numbers~\cite{davidson2015turbulence}, but it has two very unique features:
First, it arises in the low Reynolds number regime of bacterial swimming (Stokes flow). Second, in contrast to inertial turbulence, it does not exhibit a spectrum of length scales but displays one characteristic vortex size controlled by the microswimmer details~\cite{aranson2012physical,wensink2012meso,ilkanaiv2017effect}.

From the theoretical side, the main features of mesoscale turbulence have been successfully reproduced by a phenomenological continuum theory for the effective microswimmer velocity ~\cite{wensink2012meso,dunkel2013minimal,dunkel2013fluid,slomka2015generalized,oza2016generalized,bratanov2015new,james2018vortex,james2018turbulence}.
More recently, mesoscale turbulence and vortex lattices were also reported in a model of self-propelled particles with competing alignment interactions~\cite{grossmann2014vortex,grossmann2015pattern}.
Going beyond the purely phenomenological approach, we have recently presented a derivation of a hydrodynamic theory starting from Langevin equations for a generic microswimmer model~\cite{heidenreich2016hydrodynamic,reinken2018derivation}.
This theory consists of a dynamic equation for the polar order parameter field coupled to the solvent flow, the latter determined by a modified Stokes equation.
In the limit of weak coupling between orientational order and solvent flow, our hydrodynamic theory reduces to the phenomenological model.
Here, we extend the theory~\cite{reinken2018derivation} towards the effect of an orienting external field, the aim being to describe situations where swimmers are subject to an externally applied torque (stemming, e.g., from a magnetic or gravitational field).
To this end, we incorporate the field in the Langevin model and derive the additional terms in the dynamic equation for the polar order parameter.
This is outlined in section \ref{sec: theory} (and \ref{app: derivation}) of the paper.

The remainder of the paper separates into two main parts.
In the first part, we investigate the impact of an orienting external field in the \textit{full} model consisting of the polar order parameter dynamics and an explicit equation for the solvent flow (i.e., a modified Stokes equation).
Corresponding numerical results are presented in section \ref{sec: numerical}.
We show that, at intermediate field strengths, the emerging patterns become strongly asymmetric as a consequence of the externally broken symmetry.
Performing a linear stability analysis (section \ref{sec: analytical}), we find that this is due to the suppression of modes that are perpendicular to the field.
For even higher external fields, the mesoscale-turbulent state is completely suppressed and we observe a homogeneous stationary polar state.
The analytical results are then used to construct a state diagram of the system.

In the second part of the paper (section \ref{sec: reduced model}), we switch to a reduced model equivalent to the phenomenological theory, where the only dynamical variable is the effective microswimmer velocity field.
The reduced model is significantly easier to handle analytically, but still exhibits the emergence of asymmetric patterns. 
Using the framework of weakly nonlinear analysis, we investigate the underlying transition from a square vortex lattice to a traveling stripe pattern that occurs upon an increase of the external field.
Finally, we present conclusions and an outlook in section \ref{sec: conclusions}.
The paper is supplemented by seven appendices providing technical details of the calculation.

\section{Hydrodynamic theory}
\label{sec: theory}

Recently, a phenomenological model~\cite{wensink2012meso,dunkel2013minimal,dunkel2013fluid,slomka2015generalized,oza2016generalized,bratanov2015new,james2018vortex,james2018turbulence} has been proposed that reproduces the main features of mesoscale turbulence including the emergence of a characteristic length scale, i.e., vortex size~\cite{zhang2009swarming}.
It is a fourth-order field theory for the divergence-free collective microswimmer velocity and combines the Toner-Tu equation~\cite{toner1998flocks} with Swift-Hohenberg-like pattern formation. 
The main feature of the Toner-Tu equation is the transition from a disordered to a polar state corresponding to collective movement of the swimmers.
Higher order gradient terms, as introduced in the Swift-Hohenberg equation, lead to a finite wavelength instability that is responsible for the collective length scale.

In earlier publications~\cite{heidenreich2016hydrodynamic,reinken2018derivation} we have shown that an equivalent hydrodynamic theory for the microswimmer velocity can be derived via a Fokker-Planck equation approach starting from a generic Langevin model (similar to~\cite{saintillan2008instabilities,saintillan2009active}) for a system of microswimmers of length $\ell$, diameter $d$ and self-swimming speed $v_0$ with constant density $\rho$ (see~\ref{app: derivation} for a summary of the derivation).
The Langevin model includes two types of interactions:
First, there are short-range contributions stemming from an activity-driven polar interaction characterized by strength $\gamma_0$ and range $r$~\cite{hoell2018particle}. 
Second, far-field hydrodynamic effects are included via a coupling to the solvent flow field $\mathbf{u}$.
The latter is determined via an averaged Stokes equation supplemented by an appropriate ansatz for the active stress tensor containing gradient terms of up to fourth order.
The resulting coarse-grained dynamics is then given by an equation for the polar order parameter field $\mathbf{P}$ (characterizing the swimmer's mean local orientation) coupled to the solvent flow field $\mathbf{u}$. The effective velocity of the microswimmers is calculated as the sum $v_0 \mathbf{P} + \mathbf{u}$.
In the limit of weak coupling between the solvent flow and the polar order, the dependence on $\mathbf{u}$ can be neglected and the dynamics is adequately described by one field~\cite{reinken2018derivation}.
In contrast to the phenomenological approach, the coefficients of the field equation are directly linked to the parameters of the microscopic Langevin model~\cite{reinken2018derivation} (see also~\ref{app: derivation}). 

For the first part of this article, however, we will consider the \textit{full} model consisting of both the dynamics of the polar order parameter $\mathbf{P}$ and the solvent flow $\mathbf{u}$. 
The dynamical equation for $\mathbf{P}(\mathbf{x},t)$ can be conveniently written in potential form
\begin{equation}
\label{eq: P (potential form)}
\mathfrak{D}_t(\mathbf{u}) \mathbf{P} + \lambda_0 \mathbf{P}\cdot\nabla\mathbf{P} = - \frac{\delta \mathfrak{F}}{\delta\mathbf{P}}, \qquad \nabla \cdot \mathbf{P} = 0.
\end{equation}
The relaxation term is given as functional derivative of 
\begin{equation}
\label{eq: functional}
\mathfrak{F} = q(\nabla \cdot\mathbf{P})+ \frac{1}{2}\alpha |\mathbf{P}|^2 + \frac{1}{4}\beta |\mathbf{P}|^4 + \frac{1}{2}\Gamma_2 (\nabla\mathbf{P})^2 + \frac{1}{2}\Gamma_4 (\nabla\nabla\mathbf{P})^2.
\end{equation}
The derived equations~(\ref{eq: P (potential form)}) and (\ref{eq: functional}) exhibit the same features as the phenomenological model:
The isotropic--polar transition occurs when $\alpha$ changes sign from positive to negative.
The coefficient $\beta$ of the cubic term determines the saturated value of the polar solution $\sqrt{-\alpha/\beta}$.
For sufficiently strong activity $\Gamma_2$ becomes negative, which leads to a finite-wavelength instability of the homogeneous state, yielding a typical length scale of $\Lambda = 2\pi\sqrt{-2\Gamma_4/\Gamma_2}$.
Turbulent dynamics is introduced to the model via the nonlinear advection term on the left-hand side of equation~(\ref{eq: P (potential form)}), with $\lambda_0$ giving the strength of the advection term.
Finally, $q(\mathbf{x})$ is a local Lagrange multiplier enforcing the incompressibility condition $\nabla\cdot\mathbf{P} = 0$.
The assumption of incompressibility is well suited for dense suspensions where density fluctuations become small~\cite{wensink2012meso}.

In contrast to the phenomenological model, the solvent flow field $\mathbf{u}$ enters explicitly through the first term in equation~(\ref{eq: P (potential form)}). 
Its coupling to the polar order parameter field $\mathbf{P}$ is contained in the generalized derivative
\begin{equation}
\label{eq: convective derivative}
\mathfrak{D}_t(\mathbf{u}) \mathbf{P} = \partial_t \mathbf{P} + \mathbf{u} \cdot \nabla \mathbf{P} - \mathbf{\Omega} \cdot \mathbf{P} - \kappa \mathbf{\Sigma} \cdot \mathbf{P},
\end{equation}
where the vorticity tensor and the deformation rate are given by $\mathbf{\Omega} = \frac{1}{2}\big[ (\nabla \mathbf{u})^\mathrm{T} - (\nabla \mathbf{u}) \big]$ and $\mathbf{\Sigma} = \frac{1}{2}\big[ (\nabla \mathbf{u})^\mathrm{T} + (\nabla \mathbf{u}) \big]$, respectively. 
The modified Stokes equation (see~\cite{reinken2018derivation}) that determines the solvent flow field $\mathbf{u}$ reads
\begin{equation}
\label{eq: Stokes}
\nabla^2 \mathbf{u} = c_\mathrm{F} \bigg( 6 c_\mathrm{I}\mathbf{P} \cdot \nabla\mathbf{P} + \nabla^2 \mathbf{P} + \frac{1}{28}\nabla^4\mathbf{P} \bigg) + \nabla p,
\end{equation}
where $p$ is an effective pressure, and the gradient terms of $\mathbf{P}$ stem from an expansion of the active stress.

Equations~(\ref{eq: P (potential form)}) to (\ref{eq: Stokes}) are rescaled using the microswimmer length $\ell$ as length scale, the self-swimming speed $v_0$ as characteristic velocity and $\ell/v_0$ as time scale~\cite{reinken2018derivation}. 
The coefficients can then be written as
\begin{equation}
	\label{eq: coefficients}
	\eqalign{
	\lambda_0 = c_\mathrm{I}( 3 + 2a_0 P_\mathrm{r} c_\mathrm{F})/5, \qquad \kappa = a_0(3 - c_\mathrm{I})/5,\\
	\alpha = \left(1 - c_\mathrm{I}\right)/P_\mathrm{r}, \qquad \beta = \frac{3}{5}c_\mathrm{I}^2/P_\mathrm{r},\\
	\Gamma_2 = \frac{1}{10}\left(\frac{r}{\ell}\right)^2 c_\mathrm{I}/P_\mathrm{r} -\frac{a_0}{15}P_\mathrm{r}c_\mathrm{F},
	\qquad \Gamma_4 = - \frac{a_0}{420}P_\mathrm{r}c_\mathrm{F},}
\end{equation}
where $P_\mathrm{r}$, $c_\mathrm{I}$, $r/\ell$, $c_\mathrm{F}$ and $a_0$ are dimensionless parameters.
The activity is quantified by the persistence number $P_\mathrm{r} = v_0 \tau/\ell$, giving the swimming speed compared to the reorientation time $\tau$. 
The strength and range of the polar interaction are given by $c_\mathrm{I}$ and $r/\ell$, respectively. For $c_\mathrm{I} < 1$ the system favors a disordered, isotropic state, while for $c_\mathrm{I} > 1$ it favors an ordered, polar state. 
The coupling to the solvent flow is characterized by the coefficient $c_\mathrm{F}$. For the dependence of $c_\mathrm{I}$ and $c_\mathrm{F}$ on microscopic parameters of the microswimmer model, see~\ref{app: derivation}. 
Finally, flow aligning effects are characterized by the shape parameter $a_0$ which depends on the swimmer aspect ratio [see equation~(\ref{eq: shape parameter}) in \ref{app: derivation}]. 

In the present work, we extend equation~(\ref{eq: P (potential form)}) by terms incorporating an external field that affects the swimmer's orientations.
On the microscopic level, we assume that the external field generates a potential for every swimmer which depends on the angle between the field's direction $\mathbf{h}$ and swimmer orientation $\mathbf{n}$ according to a standard ferromagnetic coupling, i.e., $\Phi_\mathrm{ext} \propto - \mathbf{n} \cdot \mathbf{h}$. This term also occurs in some passive liquids in an orienting field (e.g., ferrofluids~\cite{blums1997magnetic,rosensweig2013ferrohydrodynamics}, which are suspensions of ferromagnetic colloids in a passive solvent) and was recently considered in the context of active fluids. 
Indeed, the same ansatz has been used to describe chemotactic~\cite{taktikos2011modeling} and magnetotactic~\cite{koessel2018controlling} bacteria.
It is in principle applicable to any microswimmer suspension subjected to an aligning torque exerted by an external field, as it occurs, e.g., in magnetotaxis~\cite{bazylinski2004magnetosome,waisbord2016environment,nadkarni2013comparison,popp2014polarity}, phototaxis~\cite{garcia2013light,martin2016photofocusing} or gravitaxis~\cite{fukui1985negative,tenhagen2014gravitaxis}.
Introducing the external potential in the Langevin equations and performing the same steps as in the field-free case, one arrives at the order parameter equation (for details see \ref{app: derivation})
\begin{equation}
\label{eq: P (potential form) plus field}
\mathfrak{D}_t(\mathbf{u}) \mathbf{P} + \lambda_0 \mathbf{P}\cdot\nabla\mathbf{P} = - \frac{\delta \mathfrak{F}}{\delta\mathbf{P}} + \mathbf{g}, \qquad \nabla \cdot \mathbf{P} = 0.
\end{equation}
The additional term on the right-hand side of the evolution equation is given as 
\begin{equation}
\label{eq: external field}
\mathbf{g} = B_0\mathbf{h}\cdot\bigg[\frac{2}{3}\mathbf{I} - \frac{3}{5}c_\mathrm{I}\bigg(\mathbf{P}\mathbf{P} - \frac{\mathbf{I}(\mathbf{P}\cdot\mathbf{P})}{3}\bigg) - \frac{2}{15} a_0 P_\mathrm{r} \mathbf{\Sigma}\bigg],
\end{equation}
where $B_0$ denotes the dimensionless external field strength.
The first term on the right-hand side of equation~(\ref{eq: external field}) increases the polar order in the direction of the external field, similar to what happens in passive fluids.
The second term arises due to the conservation of the unit vector $\mathbf{n}$, i.e., the microswimmer's orientation.
It leads to a saturation of $\mathbf{P}$ with increasing $B_0$, as we will later see (compare figure~\ref{fig: homogeneous stationary solution}).
Finally, the third term in equation~(\ref{eq: external field}) arises as a consequence of the closure scheme applied for the nematic order parameter tensor $\mathbf{Q}$ (see \ref{app: derivation}).
Clearly, this term incorporates a coupling to the solvent flow field.
Physically, it adds a reduction of polar order due to the flows generated by the swimmers.

\section{Numerical observations}
\label{sec: numerical}

\begin{figure}
	\raggedleft
	\includegraphics[width=0.88\linewidth]{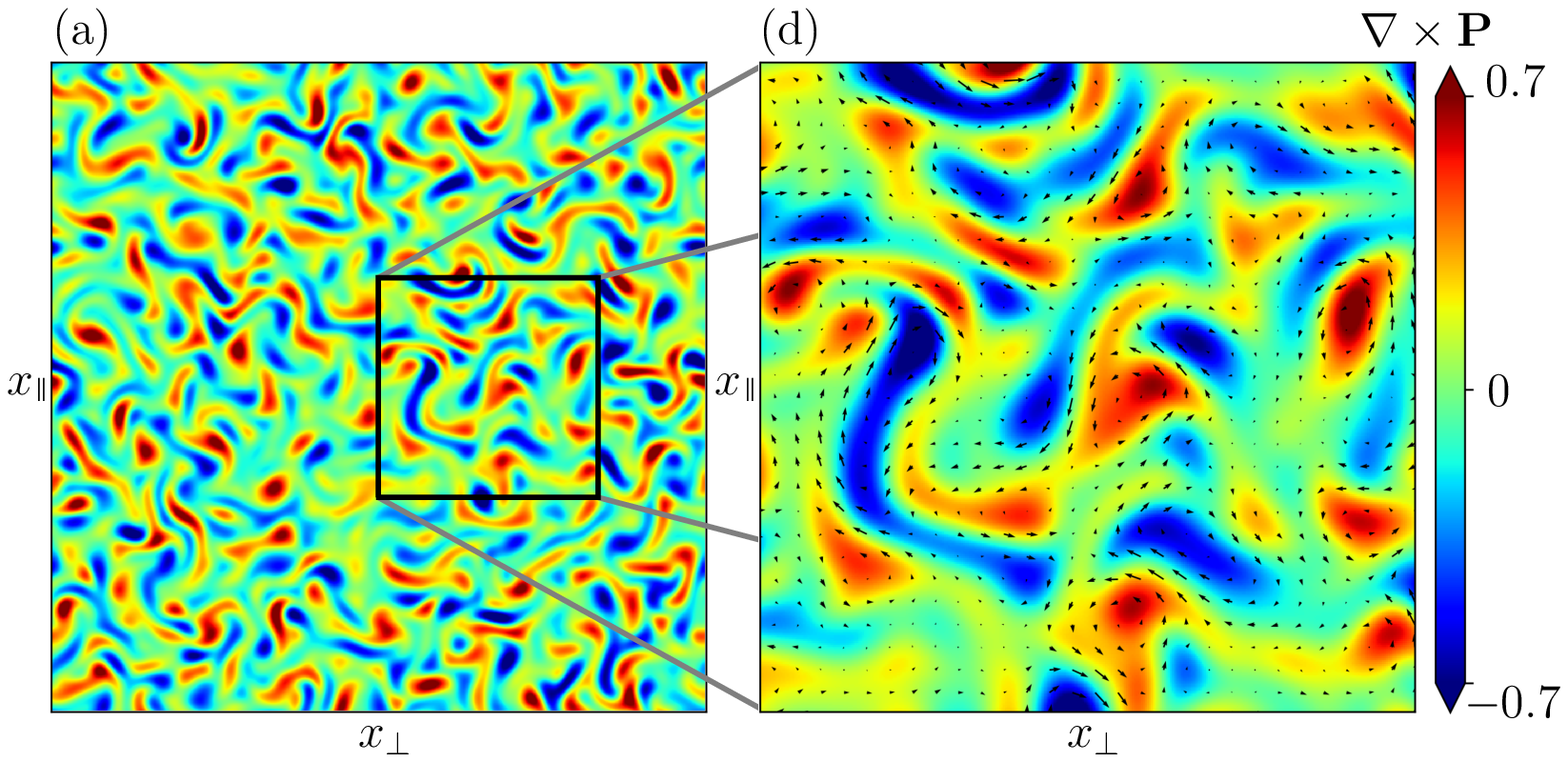}\\[0.2\baselineskip]
	\includegraphics[width=0.88\linewidth]{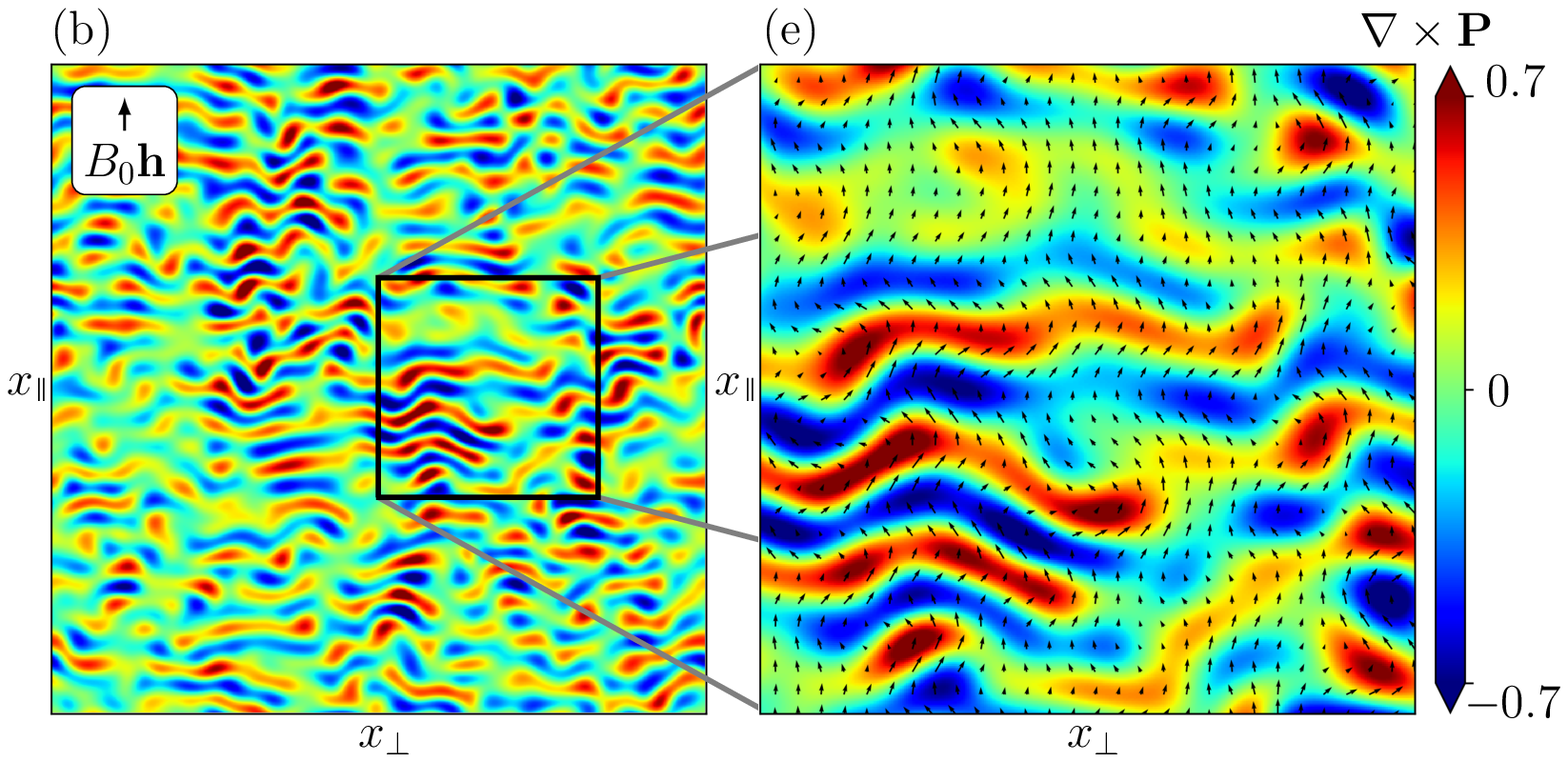}\\[0.2\baselineskip]
	\includegraphics[width=0.88\linewidth]{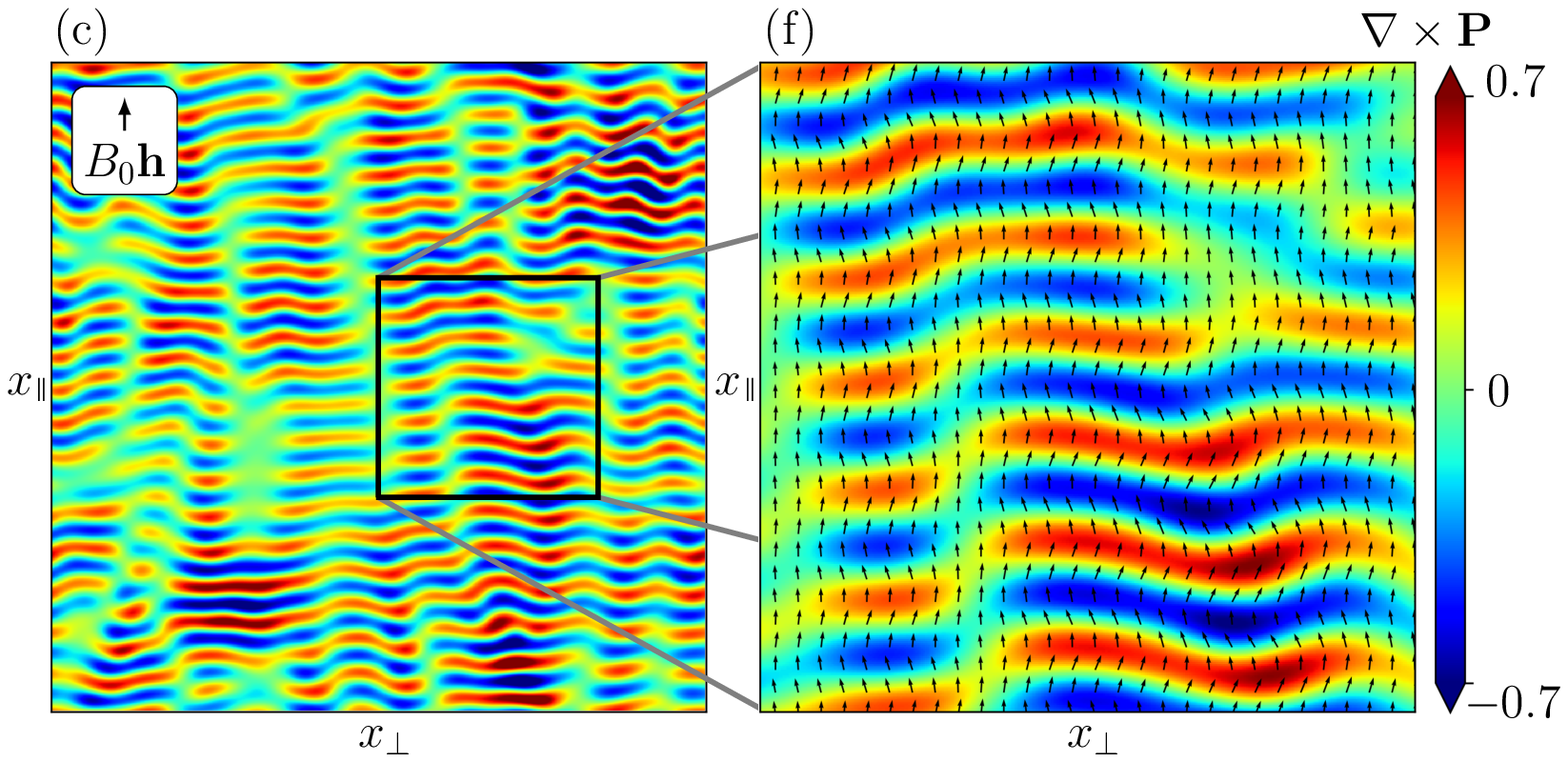}
	\caption{Snapshot of the (scalar) vorticity of the polar order parameter field as heatmap plot for (a) $B_0 = 0$, (b) $B_0 = 0.6$ and (c) $B_0 = 0.8$. Blue means clockwise, red counter-clockwise rotation. The enlarged versions (d),(e) and (f) visualize additionally the polar order parameter field by arrows, with the length corresponding to the magnitude of the field. The remaining parameters are $P_\mathrm{r} = 8$, $c_\mathrm{I} = 0.5$, $c_\mathrm{F} = 0.1$, $r/\ell = 1$ and $\ell/d = 6$.}
	\label{fig: snapshots}
\end{figure}

The aim of the present study is to explore the impact of a spatially homogeneous stationary field on the mesoscale-turbulent state observed in the absence of a field.
As a starting point, we will discuss the dynamical behavior that can be observed based on numerical solution of equation~(\ref{eq: P (potential form) plus field}) for the polar order parameter field $\mathbf{P}$, coupled to the Stokes equation~(\ref{eq: Stokes}) for the flow field $\mathbf{u}$.
The solution is performed in two-dimensional space (for the numerical methods, see~\ref{app: numerical methods}).

We start with the field-free case, $B_0 = 0$.
Here, we focus on parameters where the system is in a mesoscale turbulent state.
In figure~\ref{fig: snapshots}(a) a snapshot of the (scalar) vorticity of the polar order parameter field is shown.
For a more descriptive visualization of the dynamics in the absence of a field see the supplementary movie 1.
For better visibility, a section of the field is presented in a larger version in figure~\ref{fig: snapshots}(d).
Here, we also visualize the (vectorial) order parameter field by arrows, with the arrow length indicating the magnitude of the polar order.
The observed dynamical state is characterized by the formation, motion and decay of clockwise (blue) and counter-clockwise (red) rotating vortices.
The emerging patterns are dominated by one characteristic length or vortex size depending on the microscopic parameters of the swimmers~\cite{aranson2012physical,wensink2012meso,ilkanaiv2017effect,heidenreich2016hydrodynamic,reinken2018derivation}, hence the name mesoscale turbulence. 
This stands in contrast to inertial turbulence observed in the Navier--Stokes equation where one finds a broad spectrum of vortex sizes~\cite{davidson2015turbulence}.
For a more detailed discussion on the turbulent state without the influence of external fields, see~\cite{wensink2012meso,bratanov2015new,oza2016generalized,james2018turbulence} for the phenomenological model and~\cite{heidenreich2016hydrodynamic,reinken2018derivation} for the present model.
Note that, compared to most of the listed publications, the coefficient $\lambda_0$ is rather large and, therefore, the shape of the vortices is highly irregular.

Turning on the external field, $B_0 > 0$, introduces several new features.
First, the external field induces a net polar order in the system, that is $\langle \mathbf{P}\rangle = A^{-1}\int \rmd\mathbf{x} \mathbf{P}(\mathbf{x},t) \neq 0$ (where $A$ is the area).
The swimmer's self-propulsion speed leads to a local transport in the direction of the polar order given by $v_0\mathbf{P}$.
Thus, as a consequence of the generated net polar order, we observe a net transport in the direction of the field.
Second, the overall rotational symmetry is broken, which leads to the emergence of asymmetric patterns.
In figure~\ref{fig: snapshots}(b) and (c) this is illustrated by snapshots of the vorticity at finite (nonzero) field strengths.
In contrast to the case $B_0 = 0$, we here observe the formation of elongated structures in the vorticity field, or, more precisely, a highly irregular stripe pattern with numerous defects.
In figure \ref{fig: snapshots}(c), where the field strength is larger compared to (b), the number of defects is smaller and the stripe pattern more regular.
Interestingly, the defects are elongated in the direction parallel to the field.
The enlarged sections in figure~\ref{fig: snapshots}(e) and (f) additionally show the polar order in the field's direction.
For a visualization of the transport of the patterns, see the supplementary movies 2 and 3.
They show the dynamics for the same values of the field strength as represented in figure~\ref{fig: snapshots}(b) and (c), i.e. $B_0 = 0.6$ and $B_0 = 0.8$, respectively.
In the remainder of this work we will elucidate the observed dynamical features using analytical methods, particularly linear stability analysis and weakly nonlinear analysis.

\section{Analytical construction of the state diagram}
\label{sec: analytical}

\subsection{Homogeneous stationary solution}
\label{sec: homogeneous stationary solution}

If the activity is sufficiently weak or the external field sufficiently strong, we numerically observe a homogeneous stationary state.
This state is the starting point of our linear stability analysis.

\begin{figure}
\centering
\includegraphics[width=0.70\linewidth]{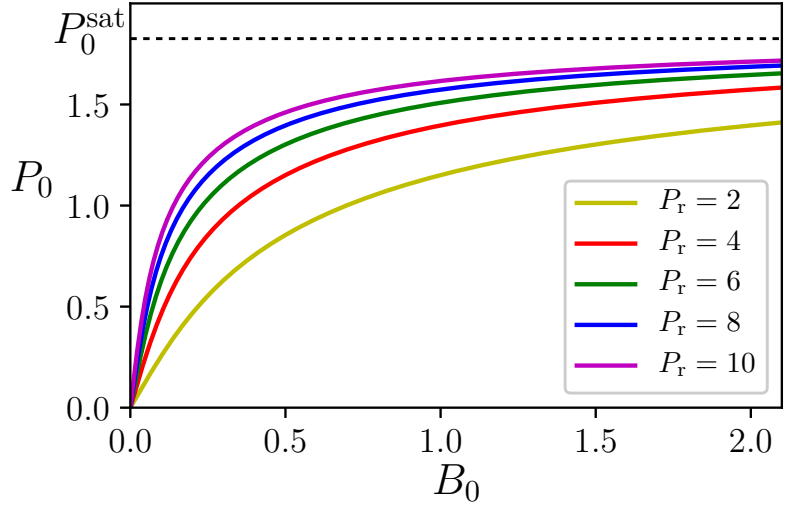}
\caption{Homogeneous stationary solution $P_0$ as function of the external field strength $B_0$ for different values of the persistence number $P_\mathrm{r}$. The saturation value $P_0^\mathrm{sat}$ appearing for very large external field strength is independent of $P_\mathrm{r}$. The remaining parameter is $c_\mathrm{I} = 0.5$.}
\label{fig: homogeneous stationary solution}
\end{figure}

Clearly, the external field breaks the rotational symmetry of the system. 
Thus, it is useful to distinguish between components of the polarization parallel and perpendicular to the field, i.e., $\mathbf{P} = (P_\para,P_\perp)$.
To calculate the homogeneous stationary solution, we assume a quiescent state, where the solvent velocity field vanishes, i.e., $\mathbf{u} = 0$.
Further, the pressure $p = p_0$ and Lagrange-multiplier $q = q_0$ are set constant in space and time.
Then, equation~(\ref{eq: P (potential form) plus field}) reduces to
\begin{equation}
\label{eq: homogeneous stationary solution}
0 = - \alpha P_0 - \beta P_0^3 + \frac{2}{3} B_0 - \frac{2}{5} B_0 c_\mathrm{I} P_0^2.
\end{equation}
The real positive solution of equation~(\ref{eq: homogeneous stationary solution}) defines the homogeneous stationary state $\mathbf{P} = (P_0,0)$.
This solution $P_0$ is plotted versus the external field strength $B_0$ in figure~\ref{fig: homogeneous stationary solution}.
As expected, the polar order grows upon increasing the field strength. 
Dividing by $B_0$ and evaluating the limit $B_0 \to \infty$, equation~(\ref{eq: homogeneous stationary solution}) simplifies to
\begin{equation}
\label{eq: homogeneous stationary solution limit}
0 = \frac{2}{3} - \frac{2}{5} c_\mathrm{I} P_0^2.
\end{equation}
The solution of equation~(\ref{eq: homogeneous stationary solution limit}) defines the saturation value $P_0^\mathrm{sat} = \sqrt{5/(3c_\mathrm{I})}$ that is approached for $B_0 \to \infty$.

\subsection{Linear stability analysis}
\label{sec: linear stability analisys}

In order to determine the linear stability of the homogeneous stationary solution $\mathbf{P} = (P_0,0)$, $q = q_0$, $\mathbf{u} = (0,0)$, $p = p_0$, we consider small perturbations, i.e.,
\begin{eqnarray}
\eqalign{
P_\para = P_0  + \delta P_\para, \qquad &P_\perp = \delta P_\perp, \qquad q = q_0 + \delta q,\\
u_\para = \delta u_\para, \qquad &u_\perp = \delta u_\perp, \qquad p = p_0 + \delta p.}
\label{eq: add perturbations}
\end{eqnarray}
For all perturbations, we make the ansatz
\begin{eqnarray}
(\delta P_\para,\delta P_\perp,\delta q,\delta u_\para,\delta u_\perp,\delta p) &= (\delta \hat{P}_\para,\delta \hat{P}_\perp,\delta \hat{q},\delta \hat{u}_\para,\delta \hat{u}_\perp,\delta \hat{p}) \ \rme^{\sigma t + \rmi \mathbf{k}\cdot \mathbf{x}},
\label{eq: perturbations form}
\end{eqnarray}
with wavevector $\mathbf{k}=(k_\para,k_\perp)$. 
We now insert equations~(\ref{eq: add perturbations}) and (\ref{eq: perturbations form}) into equations~(\ref{eq: functional}), (\ref{eq: convective derivative}), (\ref{eq: Stokes}), (\ref{eq: P (potential form) plus field}) and (\ref{eq: external field}) and linearize with respect to the perturbations.
As shown in \ref{app: linear stability analysis}, the perturbations of the velocity field $\delta\mathbf{u}$, the pressure $\delta p$ and the Lagrange multiplier $\delta q$ can be related to the perturbations of the polar order parameter $\delta\mathbf{P}$.
As a result (see \ref{app: linear stability analysis}) one obtains a linearized system involving only $\delta\mathbf{P}$, that is
\begin{equation}
\label{eq: linearized system}
\sigma \delta \hat{\mathbf{P}} = \mathbf{\Pi}(\mathbf{k})\cdot \mathbf{M}(\mathbf{k}) \cdot \delta \hat{\mathbf{P}},
\end{equation}
where the projector $\mathbf{\Pi}(\mathbf{k}) = \mathbf{I} - \mathbf{k}\mathbf{k}/|\mathbf{k}|^2$ arises as a consequence of the incompressibility of the order parameter field, and the $2\times2$-matrix $\mathbf{M}(\mathbf{k})$ is the Jacobian.
The components of $\mathbf{M}(\mathbf{k})$ are given in equation~(\ref{eq: LSA linearized system matrix}) in \ref{app: linear stability analysis}.
We obtain the complex growth rate $\sigma = \sigma_\mathrm{Re} + \rmi \sigma_\mathrm{Im}$ as a function of the wavevector $\mathbf{k}$ by calculating the eigenvalues of the matrix $\mathbf{\Pi}(\mathbf{k})\cdot \mathbf{M}(\mathbf{k})$.
The real part $\sigma_\mathrm{Re}(\mathbf{k})$, which determines the actual growth of a mode $\mathbf{k}$, is given by
\begin{eqnarray}
\label{eq: growth rate real}
\sigma_\mathrm{Re}(\mathbf{k}) = &- \alpha - \Gamma_2 |\mathbf{k}|^2 - \Gamma_4 |\mathbf{k}|^4 - \beta P_0^2 \Big(k_\para^2 + 3 k_\perp^2\Big)/|\mathbf{k}|^2 \\
&+ 3 c_\mathrm{F}c_\mathrm{I} P_0 \Big(P_0(1+\kappa) - \frac{2}{15}a_0 P_\mathrm{r} B_0\Big) k_\para^2/|\mathbf{k}|^2 \nonumber\\
&- \frac{c_\mathrm{I} }{5}B_0 P_0 \Big(3 k_\para^2 + 4 k_\perp^2\Big)/|\mathbf{k}|^2.\nonumber
\end{eqnarray}
The imaginary part $\sigma_\mathrm{Im}(\mathbf{k})$, which determines the linear traveling speed $c_0^\para$, is given by
\begin{equation}
\label{eq: growth rate imaginary}
\eqalign{
\sigma_\mathrm{Im}(\mathbf{k}) &= \Big[- \lambda_0 P_0 + \frac{c_\mathrm{F}}{2} \Big(P_0(1+\kappa) - \frac{2}{15}B_0 a_0 P_\mathrm{r}\Big)\Big(1-\frac{|\mathbf{k}|^2}{28}\Big)\Big] k_\para \\
&= - c_0^\para  k_\para.
}
\end{equation}
It is seen that $\sigma_\mathrm{Re}(\mathbf{k})$ depends not only on the magnitude $|\mathbf{k}| = k$ of the wavevector but also on its direction.
For $B_0 = 0$ and $c_\mathrm{F} = 0$, one reproduces the growth rate obtained in~\cite{dunkel2013minimal}.

\subsection{State diagram}
\label{sec: state diagram}

\begin{figure}
\raggedleft
\includegraphics[width=0.85\linewidth]{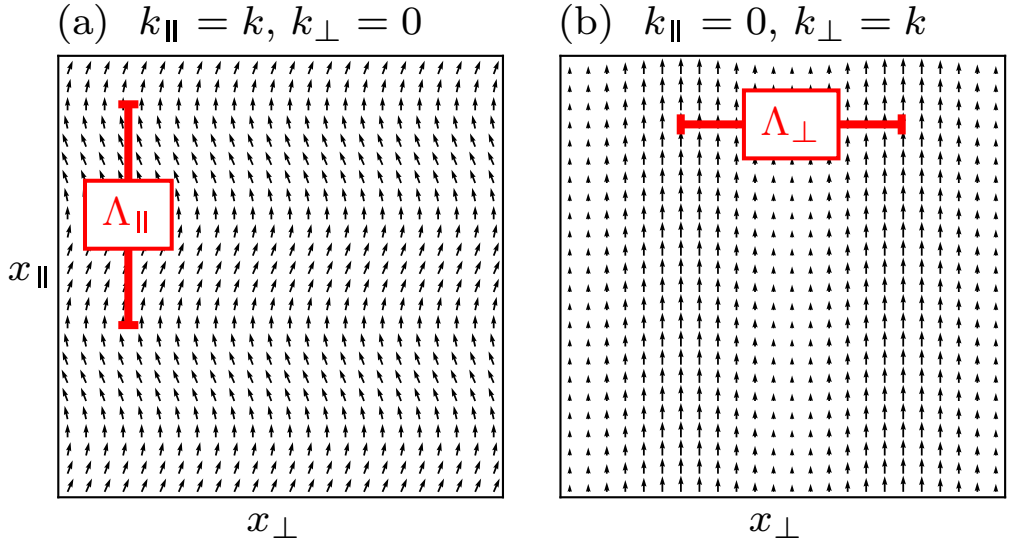}
\caption{Form of the perturbed polar order parameter field in the two limiting cases of a wavevector (a) parallel and (b) perpendicular to the external field. The corresponding dominating wavelength $\Lambda_\para$ or $\Lambda_\perp$, respectively, is indicated in red}
\label{fig: perturbations}
\end{figure}

Due to the explicit dependence of the growth rate $\sigma_\mathrm{Re}(\mathbf{k})$ on the wavevector's direction, it makes sense to distinguish between two limiting cases illustrated in figure~\ref{fig: perturbations}: perturbations with a wavevector that is purely parallel to the external field, i.e., $k_\para = k$, $k_\perp = 0$, or purely perpendicular to the field, i.e., $k_\para = 0$, $k_\perp = k$.
The incompressibility condition of the order parameter field, i.e., $k_\para \delta P_\para + k_\perp \delta P_\perp = 0$, then dictates the form of the respective perturbations.
For a parallel wavevector, the component $\delta P_\para$ must vanish.
Therefore, the linearly unstable pattern is a perturbation in the perpendicular direction with periodicity in the parallel direction [see figure~\ref{fig: perturbations}(a)].
The corresponding wavelength is given by $\Lambda_\para = 2\pi/k$.
This type of pattern manifests itself as stripes in the vorticity of the field.
Analogously, for a perpendicular wavevector, the component $\delta P_\perp$ must vanish and the pattern is a perturbation along the field with periodicity in the perpendicular direction ($\Lambda_\perp = 2\pi/k$) [see figure~\ref{fig: perturbations}(b)].
The occurrence of both, modes in the parallel and in the perpendicular direction, results in a square vortex lattice.

\begin{figure}
\raggedleft
\includegraphics[width=0.85\linewidth]{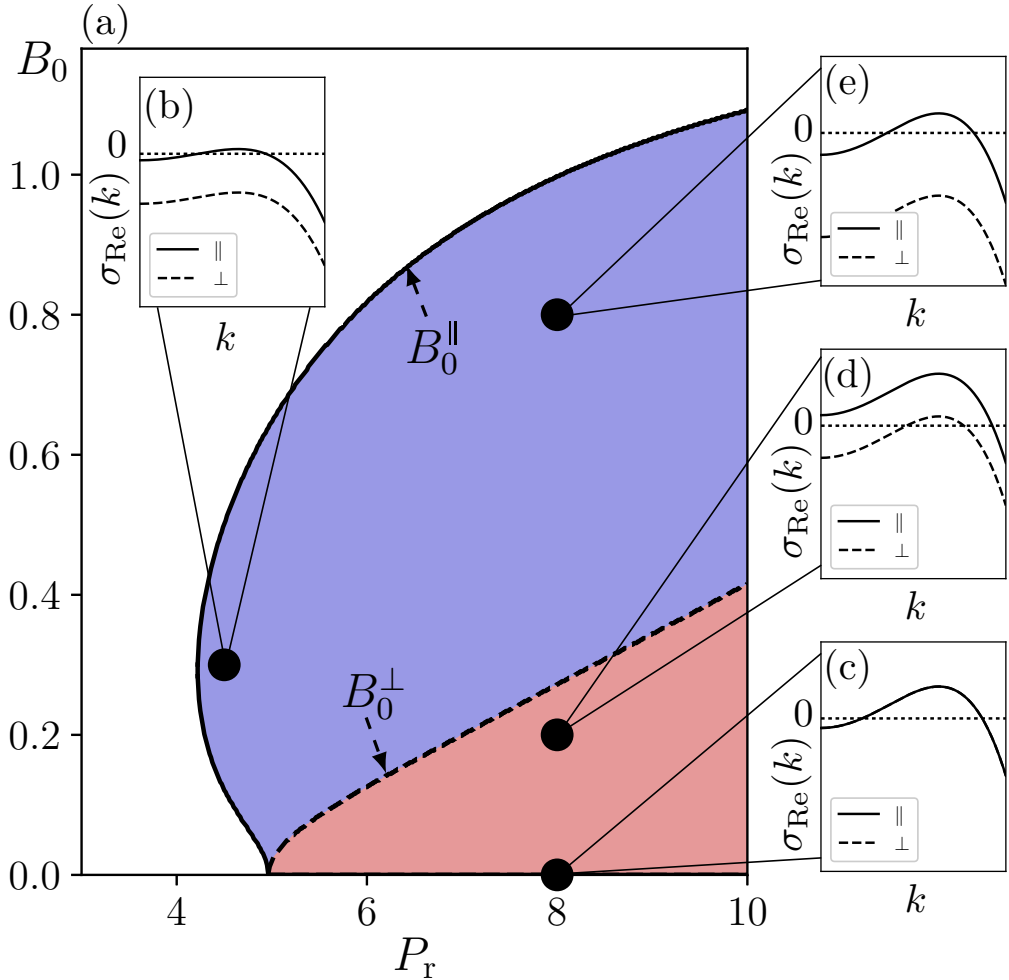}
\caption{(a) State diagram in the $P_\mathrm{r}$-$B_0$-plane obtained by linear stability analysis. (b) - (e) Real part of the growth rate as function of the wavenumber for $\mathbf{k}\para\mathbf{h}$ and $\mathbf{k}\perp\mathbf{h}$. The solid line in (a) denotes the field strength where parallel modes start to grow, the dashed line where perpendicular modes start to grow. The remaining parameters are $c_\mathrm{I} = 0.5$, $r/\ell = 1$, $c_\mathrm{F} = 0.1$ and $\ell/d = 6$. Note that the diagram is independent of the magnitude of the nonlinear advection term $\lambda_0$.}
\label{fig: state diagram}
\end{figure}

The results obtained from linear stability analysis and the distinction between the two limiting cases for the perturbation enable the construction of a state diagram in the plane spanned by $P_\mathrm{r}$ and $B_0$, see figure~\ref{fig: state diagram}(a).
The diagram is supplemented by plots of the growth rate as function of the wavevector, see figures~\ref{fig: state diagram}(b) to (e).
Regions where the homogeneous stationary state described in section \ref{sec: homogeneous stationary solution} is stable are indicated by a white background color.
Considering first the case $B_0 = 0$, we observe an instability at a critical value ($P_\mathrm{r} \approx 5$) where mesoscale turbulence sets in.
This is reflected by a band of unstable modes (see figure~\ref{fig: state diagram}(c)), one of which grows the fastest and yields a typical vortex size $\Lambda = 2\pi\sqrt{-2\Gamma_4/\Gamma_2}$.
Note that, without an external field, the full rotational symmetry is still intact and the two curves for parallel and perpendicular wavevector perturbations coincide.
For a (numerical) snapshot of the order parameter field $\mathbf{P}$ and its vorticity $\nabla\times\mathbf{P}$ in the mesoscale turbulent state at $B_0 = 0$, see figure~\ref{fig: snapshots}(a).

Keeping the persistence number at a constant value within the zero-field turbulent state, e.g., $P_\mathrm{r} = 8$, and increasing the external field from zero results in a shift of the growth rate curve $\sigma_\mathrm{R}$ depending on the wavevector's direction (see figure~\ref{fig: state diagram}(d)).
It is seen that perturbations with perpendicular wavevector become suppressed above a field strength of $B_0^\perp(P_\mathrm{r})$, see dashed line in figure~\ref{fig: state diagram}(a).
In contrast, perturbations with parallel wavevector still grow until a larger field strength of $B_0^\para(P_\mathrm{r})$ is reached, see solid line in figure~\ref{fig: state diagram}(a).
Between the two ``critical'' field strengths $B_0^\para$ and $B_0^\perp$, the growth of parallel modes leads to asymmetric patterns which are elongated in the perpendicular direction.
This is consistent with our numerical observations shown in figure~\ref{fig: snapshots}(b) and (c).
Interestingly, the characteristic length given by the critical mode $k_\mathrm{c}$, i.e., the maximum of $\sigma_\mathrm{Re}(\mathbf{k})$, is not influenced by the external field [compare figure~(\ref{fig: state diagram})(c) to (e)].

Moreover, we observe a very intriguing feature in the case of a perturbation with $\mathbf{k}\para\mathbf{h}$.
Here, the growth rate curve is first shifted upwards for intermediate field strengths before being shifted down for stronger fields.
This is due to the explicit coupling of the solvent flow to the generated polar order in the system.
As discussed in a variety of publications~\cite{Marchetti2013,simha2002hydrodynamic,ramaswamy2010mechanics}, active suspensions with uniform orientational order such as active nematics~\cite{edwards2009spontaneous} are intrinsically unstable.
The interplay between actively generated flow and the resulting flow alignment of the elongated particles destabilizes the uniaxial order.
In active nematics, this results in a bend instability for particles that generate an extensile active stress and a splay instability for particles that generate a contractile active stress~\cite{ramaswamy2010mechanics}.
In our model of polar microswimmers, a similar mechanism leads to the upwards shift of the growth rate curve of parallel modes for intermediate field strengths.
The external field generates orientational order, which in turn induces a solvent flow that destabilizes the homogeneous state due to the flow alignment of the swimmers.
For higher field strengths, however, this interplay between polar order and solvent flow is outweighed by the other terms in the growth rate [equation~\ref{eq: growth rate real}] that suppress the instability.
This behavior is reflected in the state diagram in figure~\ref{fig: state diagram}(b) by the pronounced ``bulge'' for persistence numbers below $P_\mathrm{r} \approx 5$.

Finally, note that the growth rate $\sigma_\mathrm{Re}$ [see equation~(\ref{eq: growth rate real})] does not depend on the nonlinear advection term in equation~(\ref{eq: P (potential form)}).
The imaginary part $\sigma_\mathrm{Im}$, however, is strongly dependent on to the parameter $\lambda_0$ characterizing the magnitude of the advection term [see equation~(\ref{eq: growth rate imaginary})].
Thus, in the framework of linear stability analysis, the influence of the nonlinear advection term is restricted to transporting the emerging patterns with a traveling speed of $c_0^\para$, given by equation~(\ref{eq: growth rate imaginary}), in the direction of the external field.
Comparing the state diagram in figure~\ref{fig: state diagram} to the full numerical solution of equations~(\ref{eq: P (potential form) plus field}) and (\ref{eq: Stokes}), we find that the analytical calculations are consistent with the numerical results.
They accurately predict the onset of the mesoscale-turbulent state, the complete suppression of the instability for high field strength and the region of asymmetric patterns for intermediate field strength (compare figure~\ref{fig: snapshots}).

\section{Reduced model}
\label{sec: reduced model}

\begin{figure}
	\centering
	\includegraphics[width=0.65\linewidth]{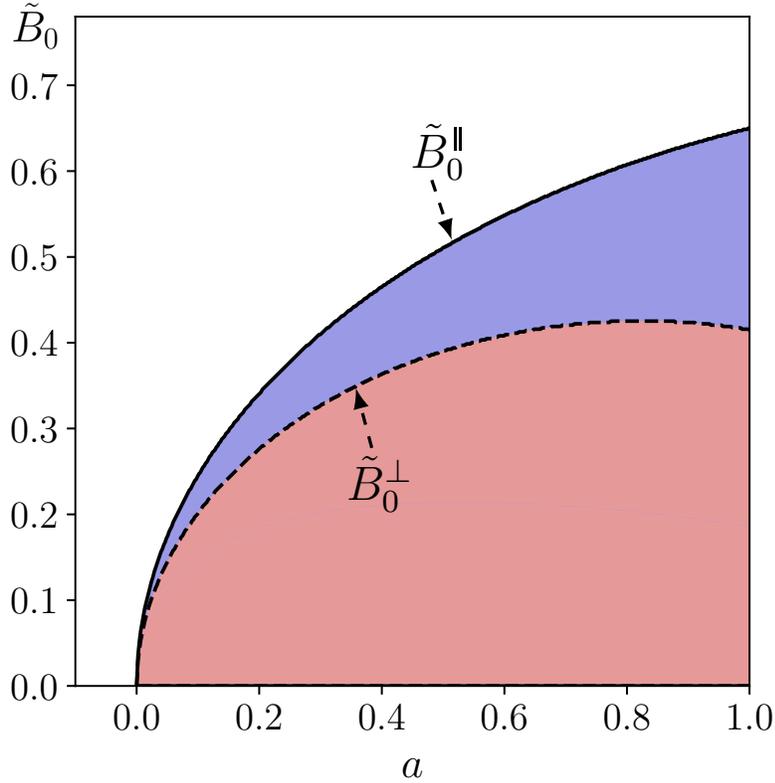}
	\caption{State diagram for the reduced model in the $a$-$\tilde{B}_0$-plane obtained by linear stability analysis. The solid line in the state diagram denotes the field strength where parallel modes start to grow, the dashed line where perpendicular modes start to grow. Compared to the state diagram of the full model (figure~\ref{fig: state diagram}) the bulge for $P_\mathrm{r} < 5$ which corresponds to $a < 0$ is absent. The coefficient of the cubic term is set to $b = 0.1$. Note that the diagram is independent of the magnitude of the nonlinear advection term $\lambda$.}
	\label{fig: state diagram reduced}
\end{figure}

In order to characterize the emergence of patterns in more detail, it seems appropriate to perform a weakly nonlinear analysis.
However, it turns out that this is an extremely difficult task when starting from the model equations discussed so far.
For the remainder of this work we thus consider a reduced model, which still gives the essential physics.

As we discussed in our previous publication~\cite{reinken2018derivation}, the response of the flow field to the forces exerted by every swimmer on the surrounding fluid scales with the coefficient $c_\mathrm{F}$. 
The latter is proportional to the ratio of active to viscous forces.
In the limit $c_\mathrm{F} \ll 1 $, the collective dynamics of the suspension depends solely on the dynamics of the polar order parameter.
Introducing an effective microswimmer velocity proportional to $\mathbf{P}$ then reproduces the phenomenological model~\cite{wensink2012meso,dunkel2013minimal,dunkel2013fluid,slomka2015generalized,oza2016generalized,bratanov2015new,james2018turbulence} briefly introduced in section~\ref{sec: theory}, with the notable difference that we can calculate all the coefficients as functions of parameters of the microswimmer model.
Note that the coefficient $\Gamma_2$ has to be negative in order to obtain a mesoscale-turbulent state.
Thus, additionally to the limit $c_\mathrm{F} \ll 1 $, the product $P_\mathrm{r} c_\mathrm{F}$ still has to be sufficiently large compared to $c_\mathrm{I}$ [see equation~\ref{eq: coefficients}].
In this work, we rescale space and time by the length and time scale of the pattern formation, i.e., the inverse of the critical mode $k_\mathrm{c} = \sqrt{-\Gamma_2/(2\Gamma_4)}$, and the inverse of the corresponding maximum of the growth rate $\sigma_\mathrm{Re}(\alpha = 0, B_0 = 0) = \Gamma_2^2/(4\Gamma_4)$.
Further, we rescale by the saturation value for the polar order parameter at $B_0 \to \infty$ and obtain the scaled collective microswimmer velocity via $\mathbf{v} = \mathbf{P}/P_0^\mathrm{sat} $, where $P_0^\mathrm{sat} = \sqrt{5/(3 c_\mathrm{I})}$.
The rescaled equation for $\mathbf{v}$ is then given by
\begin{equation}
\label{eq: dynamic equation scaled}
\eqalign{
\partial_t \mathbf{v} + \lambda \mathbf{v}\cdot\nabla\mathbf{v} = &- \nabla q + a \mathbf{v} - b |\mathbf{v}|^2\mathbf{v} - (1 + \nabla^2)^2 \mathbf{v}\\
&+ \tilde{B}_0\mathbf{h}\cdot\Big[\mathbf{I} - \frac{1}{2}\Big(3\mathbf{v}\mathbf{v} - \mathbf{I}(\mathbf{v}\cdot\mathbf{v})\Big)\Big].}
\end{equation}
The advantage of this scaling is that the number of independent coefficients is now reduced to four: 
First, the coefficient $\lambda$ determines the strength of the nonlinear advection term.
Second, the coefficient $a$ characterizes the distance to the onset of the instability at $\tilde{B}_0 = 0$.
For $a < 0$, the system favors the isotropic homogeneous state, while for $a > 0$ the system develops mesoscale turbulence.
Note that for $a > 1$, the homogeneous stationary solution becomes polar.
In this work, we will focus on the case $a < 1$.
Third, the coefficient $b$ of the cubic term in $\mathbf{v}$ is responsible for the saturation of the emerging patterns.
Fourth, the external field strength is given by $\tilde{B}_0$.
The scaling of the Lagrange multiplier $q$ enforcing the incompressibility $\nabla \cdot \mathbf{v} = 0$ is irrelevant, thus, we keep the same notation as before.
For the explicit dependencies of the four remaining coefficients $\lambda$, $a$, $b$ and $\tilde{B}_0$ on the coefficients of the full model for the dynamics of $\mathbf{P}$ given in equations (\ref{eq: P (potential form) plus field}) and (\ref{eq: external field}), see \ref{app: coefficients reduced model}.

The reduced and rescaled model is significantly easier to handle than the full version but still exhibits the emergence of asymmetric patterns due to the external field.
The homogeneous stationary solution $\mathbf{v} = (v_\para,v_\perp) = (V_0,0)$ of the reduced model is calculated via
\begin{equation}
\label{eq: homogeneous stationary solution scaled}
0 = (a - 1) V_0 - b V_0^3 + \tilde{B}_0 - \tilde{B}_0 V_0^2.
\end{equation}
Analogous to the full model, performing a linear stability analysis yields the complex growth rate with real part
\begin{equation}
\sigma_\mathrm{Re}(\mathbf{k})= a - 1 + 2 |\mathbf{k}|^2 - |\mathbf{k}|^4 - b V_0^2\frac{k_\para^2 + 3 k_\perp^2}{|\mathbf{k}|^2} - \frac{1}{2}\tilde{B}_0 V_0\frac{3 k_\para^2 + 4 k_\perp^2}{|\mathbf{k}|^2}.
\label{eq: growth rate rescaled}
\end{equation}
Similar to the corresponding function $\sigma_\mathrm{\mathbf{k}}$ of the full model [compare equation~(\ref{eq: growth rate real})], equation~(\ref{eq: growth rate rescaled}) yields a finite-wavelength instability.
Due to the present scaling, the critical mode is now given by $k_\mathrm{c} = 1$.
Further, we obtain the linear traveling speed as
\begin{equation}
c_0^\para = - \frac{\sigma_\mathrm{Im}(\mathbf{k})}{k_\para} = \lambda V_0.
\label{eq: traveling speed rescaled}
\end{equation}
Note that, for the sake of brevity, we use the same notation for the growth rate and related concepts as for the full model.

The state diagram obtained from the stability analysis of equation~(\ref{eq: dynamic equation scaled}) is shown in figure~\ref{fig: state diagram reduced}.
As in the full model we find a region where perturbations with a parallel wavevector grow but perpendicular wavevector perturbations decay.
However, compared to figure~\ref{fig: state diagram}, the bulge on the left-hand side is absent.
This is because the reduced model is solely given by equation~(\ref{eq: dynamic equation scaled}) and, thus, the coupling of the externally generated polar order to the Stokes equation~(\ref{eq: Stokes}) is neglected.
As discussed in section~\ref{sec: state diagram}, this coupling is essential for the mechanism producing the bulge.

\subsection{Weakly nonlinear analysis}
\label{sec: weakly nonlinear analysis}

\begin{figure}
\raggedleft
\includegraphics[width=1.0\linewidth]{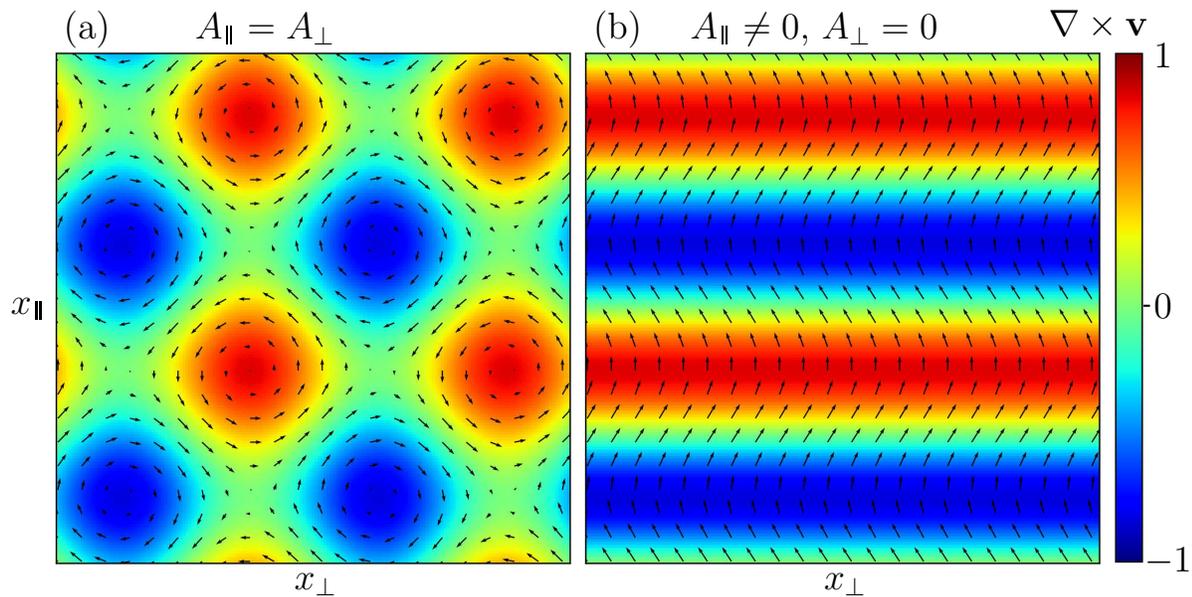}
\caption{Schematic visualization of the effective microswimmer velocity field and corresponding vorticity for (a) a regular vortex lattice and (b) a stripe pattern. The length and direction of the arrows denote the strength and direction of the field, respectively. The vorticity of the field is given as background color, where blue means clockwise, red counter-clockwise rotation.}
\label{fig: patterns}
\end{figure}

To analyze in more detail the emerging patterns in the reduced model, we now perform a weakly nonlinear analysis~\cite{cross1993pattern,newell1993order} for two different values of $B_0$:
In section \ref{sec: weakly nonlinear analysis no field} we start with $\tilde{B}_0 = 0$ (and $a > 0$), where the unstable modes have no preferred direction.
The emerging pattern in this case is a square vortex lattice.
In section \ref{sec: weakly nonlinear analysis field}, we then consider the case $\tilde{B}_0^\perp < \tilde{B}_0 < \tilde{B}_0^\para$ (and $a > 0$), where the system's rotational symmetry is broken and modes with a wavevector parallel to the field dominate the dynamical behavior.
Here, we observe the emergence of stripes stacked in the field's direction.
The effective microswimmer velocity field for these two regular patterns is visualized in figure~\ref{fig: patterns}(a) and (b), respectively.
Various technical details related to the weakly nonlinear analysis are provided in \ref{app: weakly nonlinear analysis}.

\subsubsection{Case of zero external field}
\label{sec: weakly nonlinear analysis no field}

In the absence of an external field, the onset of the instability occurs at $a = 0$ (see figure~\ref{fig: state diagram reduced}).
Due to the rotational symmetry, the weakly nonlinear analysis for this case is essentially standard (see \cite{cross1993pattern,newell1993order}).
We introduce a small parameter $\varepsilon$ characterizing the distance to the  bifurcation via the growth rate of critical modes $\varepsilon^2 = \sigma_\mathrm{Re} (B_0 = 0,k = 1)= a$ [compare equation~(\ref{eq: homogeneous stationary solution scaled}) and (\ref{eq: growth rate rescaled})].
This allows the definition of a slow time scale $T = \varepsilon^2 t$ and corresponding spatial variable $\mathbf{X} = \varepsilon \mathbf{x}$ (motivated by the fact that the growth rate scales in lowest order quadratic in $\mathbf{k}$).
These long time and space variables characterize the scales on which the amplitude of the emerging patterns evolves.
The next step is to expand the effective microswimmer velocity field $\mathbf{v}$ in orders of $\varepsilon$.
In the present case, unstable modes have no preferred direction and the emerging pattern is a square vortex lattice [see figure~\ref{fig: patterns}(a)] which is formed by critical wavenumber modes perpendicular to each other.
Although their directions are arbitrary at $B_0 = 0$, for convenience, we choose one mode parallel and one mode perpendicular to the direction of the external field which will be set to finite values in the next section.
We thus also consider the (scalar) components $v_\para$ and $v_\perp$ of the effective velocity field.
Taking into account that the incompressibility condition, i.e., $\nabla\cdot\mathbf{v} = 0$, has to be satisfied, the expansion reduces in lowest order to
\begin{equation}
\label{eq: expansion vortex lattice}
\eqalign{
v_\para(\mathbf{x},t,\mathbf{X},T) &= \varepsilon A_\perp (\mathbf{X},T) \rme^{\rmi x_\perp} + \mathrm{c.c.} \, ,\\
v_\perp(\mathbf{x},t,\mathbf{X},T) &= \varepsilon A_\para (\mathbf{X},T) \rme^{\rmi x_\para} + \mathrm{c.c.} \, ,} 
\end{equation}
where $A_\para$ and $A_\perp$ are the amplitudes of the two modes and $\mathrm{c.c.}$ denotes the complex conjugate.
In the exponential functions, we have used $k_\mathrm{c} = 1$.
Now, we insert the expansion equation~(\ref{eq: expansion vortex lattice}) into equation~(\ref{eq: dynamic equation scaled}) and use the definitions for the slow time scale $T$ and corresponding spatial variable $\mathbf{X}$.
Matching terms with the same order of $\varepsilon$ yields the amplitude equations for the parallel and perpendicular mode in $\mathcal{O}(\varepsilon^3)$,
\begin{equation}
\eqalign{
\partial_t A_\para = a A_\para - b\Big(3 |A_\para|^2 + 2 |A_\perp|^2 \Big)A_\para + 4 \partial_{x_\para}^2 A_\para,\\
\partial_t A_\perp = a A_\perp - b\Big(3 |A_\perp|^2 + 2 |A_\para|^2 \Big)A_\perp + 4 \partial_{x_\perp}^2 A_\perp,}
\label{eq: amplitude equations vortex lattice}
\end{equation}
where we already scaled back to the fast time and length scales, $t$ and $\mathbf{x}$, respectively.
It is seen that the two amplitude equations~(\ref{eq: amplitude equations vortex lattice}) correspond to two coupled Ginzburg-Landau equations~\cite{aranson2002world}.
This result was already obtained in~\cite{james2018turbulence}, where the pattern formation in the reduced model equation~(\ref{eq: dynamic equation scaled}) in the absence of an external field is investigated.
Physically, equations~(\ref{eq: amplitude equations vortex lattice}) describe a relaxation towards a uniform, stationary state characterized by the homogeneous stationary solution $A^\mathrm{sat}_\para = A^\mathrm{sat}_\perp = \sqrt{a/(5b)}$, corresponding to a regular square vortex lattice.

\subsubsection{Finite external field}
\label{sec: weakly nonlinear analysis field}

Being interested in the symmetry-broken state, we now move on to perform a weakly nonlinear analysis in the region of the state diagram (figure~\ref{fig: state diagram reduced}) where perturbations with a parallel wavevector grow but perpendicular wavevector perturbations decay. 
Starting from the homogeneous stationary state at large fields $\tilde{B}_0 > \tilde{B}_0^\parallel$, the onset of the instability (at $a > 0$) occurs at $\tilde{B}_0^\parallel (a)$, i.e., the solid line in figure~\ref{fig: state diagram reduced}.
In analogy to the analysis for $\tilde{B}_0 = 0$, we introduce a small parameter $\varepsilon$ quantifying the distance to the bifurcation.
In the present case, we define $\varepsilon$ via
\begin{equation}
\varepsilon^2 = \sigma_\mathrm{Re}^\para = \sigma_\mathrm{Re}(k_\para = 1, k_\perp = 0) =  a - b V_0^2 - \frac{3}{2}\tilde{B}_0 V_0,
\label{eq: bifurcation point}
\end{equation}
where we have used equation~(\ref{eq: growth rate rescaled}).
The slow time scale is again defined by $T = \varepsilon^2 t$.
The scaling of the corresponding spatial variable is given by $\mathbf{X} = \varepsilon (\mathbf{x} - \mathbf{v}_\mathrm{g} t)$, where we introduced the group velocity $\mathbf{v}_\mathrm{g}$ for the following reason:
In contrast to the case $\tilde{B}_0 = 0$, the system at $\tilde{B}_0 > 0$ displays a net polarization, and the emerging stripe pattern [see figure~\ref{fig: patterns}(a)] is traveling in the field's direction with the traveling speed $c_\para$.
We also expect modulations of the pattern to travel through the system. 
The corresponding velocity of the modulations is denoted as group velocity $\mathbf{v}_\mathrm{g}$ and is, at this point, still to be determined.
The next step is to expand the effective microswimmer velocity field $\mathbf{v}$ with respect to the distance to the bifurcation $\varepsilon$.
Here, we use an ansatz which incorporates only parallel modes characterized by multiples of the critical wavenumber ($k_\mathrm{c} = 1$). 
In contrast to equation~(\ref{eq: expansion vortex lattice}), perpendicular modes are not considered, since they decay in the considered region in the state diagram.
The full expansion is given as
\begin{equation}
\label{eq: expanison effective velocity full}
\eqalign{
v_\para(\mathbf{x},t,\mathbf{X},T) = V_0 + &\sum_{n=1}^{\infty} \sum_{m=0}^{n} \varepsilon^n v_{n,m}^\para (\mathbf{X},T) \rme^{\rmi m (x_\para - c_\para t)} + \mathrm{c.c.} \, ,\\
v_\perp(\mathbf{x},t,\mathbf{X},T) = &\sum_{n=1}^{\infty} \sum_{m=0}^{n} \varepsilon^n v_{n,m}^\perp (\mathbf{X},T) \rme^{\rmi m (x_\para - c_\para t)} + \mathrm{c.c.} \, ,}
\end{equation}
where $V_0$ denotes the homogeneous stationary solution [$\mathbf{v} = (v_\para,v_\perp) = (V_0,0)$] and $\mathrm{c.c.}$ the complex conjugate.
Similarly, we also expand the local Lagrange multiplier $q$ and traveling speed $c^\para$ [see equation~(\ref{eq: Lagrange multiplier expansion}) and (\ref{eq: traveling speed expansion}) in \ref{app: weakly nonlinear analysis}].
Inserting all expansions [equations~(\ref{eq: expanison effective velocity full}), (\ref{eq: traveling speed expansion}) and (\ref{eq: Lagrange multiplier expansion})] into the dynamic equation~(\ref{eq: dynamic equation scaled}) and using the incompressibility constraint $\nabla \cdot \mathbf{v} = 0$, one obtains solvability conditions in different modes $\rme^{\rmi m (x_\para - c_\para t)}$ and different orders of $\varepsilon$ that all have to be satisfied.
In $\mathcal{O}(\varepsilon^3)$ we obtain a dynamic equation for the leading order amplitude $v_{1,1}^\perp$, which we will denote as $A$ in what follows for the sake of brevity.
All other contributions in the expansion either vanish, contribute in orders higher than $\mathcal{O}(\varepsilon^3)$ or can be expressed as functions of $A$, i.e., are slaved to the dominating mode. 
For further technical details of the weakly nonlinear analysis, see \ref{app: weakly nonlinear analysis}.
The final amplitude equation for the dominating mode is 
\begin{equation}
\partial_t A +  v^\para_\mathrm{g} \partial_{x_\para} A = \sigma^\para_\mathrm{Re} A - g |A|^2 A + D_\para \partial_{x_\para}^2 A + D_\perp \partial_{x_\perp}^2 A,
\label{eq: amplitude equation stripes}
\end{equation}
where we already scaled back to the fast scales.

We find that the coefficient $g > 0$ [see equation~(\ref{eq: cubig coefficient g})], thus, the bifurcation at $\tilde{B}_0 = \tilde{B}_0^\para$ is supercritical.
As for the vortex lattice at $\tilde{B}_0 = 0$, the obtained amplitude equation~(\ref{eq: amplitude equation stripes}) corresponds to a real-valued Ginzburg--Landau equation~\cite{aranson2002world} and describes a relaxation process to a uniform amplitude, given by the homogeneous stationary solution $A^\mathrm{sat} = \sqrt{\sigma_\mathrm{Re}^\para / g}$.
However, there are several features not present in equations~(\ref{eq: amplitude equations vortex lattice}):
First, the stripe pattern and modulations of it are transported along the direction of the external field.
This transport is characterized by the speed $c^\para$ and the group velocity $v_\mathrm{g}^\para = \lambda V_0$, which is equal to the linear traveling speed $c_0^\para$.
In addition to the result from linear stability analysis, the traveling speed $c^\para$ contains contributions of higher order in $\varepsilon$ [see equation~(\ref{eq: traveling speed second order}) in \ref{app: weakly nonlinear analysis}].
Up to second order, we have
\begin{equation}
c^\para = \lambda (V_0 + D_0 |A|^2).
\label{eq: traveling speed modified}
\end{equation}
Thus, the amplitude of the emerging patterns modifies the speed with which they travel through the system.
A second unique feature of the symmetry-broken case is the anisotropic diffusion of pattern modulations, as reflected by the difference of the diffusions constants $D_\para$ and $D_\perp$, respectively,
\begin{equation}
D_\para = 4, \qquad D_\perp =  2 b V_0^2 + \frac{1}{2} \tilde{B}_0 V_0.
\label{eq: diffusion coefficients}
\end{equation}
At the parameters considered, the coefficient for parallel transport is approximately one order of magnitude higher than for perpendicular transport.
This leads to an interesting additional effect, which we discuss in detail in section~\ref{sec: conclusions}.

\subsection{Transition between square vortex lattice and stripe pattern}
\label{sec: transition stripes lattice}

Having obtained the amplitude equations~(\ref{eq: amplitude equations vortex lattice}) and (\ref{eq: amplitude equation stripes}) for the square vortex lattice and the traveling stripe pattern, respectively, there remains the question about their predictive power.
Indeed, one would expect that only situations with weak nonlinearities, i.e. situations near the onset of the respective instabilities, are adequately described in the framework of weakly nonlinear analysis~\cite{aranson2012physical}.
These weak nonlinearities include the quadratic and the cubic term in equation~(\ref{eq: dynamic equation scaled}), which are responsible for the saturation of the amplitudes, as well as the linear part of the self-advection term that is responsible for the transport of the patterns in the direction of the field.
This linear part is proportional to the net velocity $\langle \mathbf{v}\rangle$ and can be written as $\lambda\langle \mathbf{v}\rangle \cdot \nabla \mathbf{v}$ [compare equation~(\ref{eq: dynamic equation scaled})].
However, the weakly nonlinear analysis does not capture the full nonlinearity of the advection term, as it is hard to handle analytically.
From classical turbulence theory it is known that this term transfers energy that is inserted into the system between different length scales~\cite{davidson2015turbulence}.
This contradicts the basis of weakly nonlinear analysis, i.e., the assumption that the dominant mode is given by the critical wavenumber.
For example, for two-dimensional flows, the energy cascade decreases the dominant wavenumber in the system, (see also figure~\ref{fig: mode shift} in \ref{app: shift of the dominating mode} and~\cite{james2018turbulence} for a more detailed discussion).

\begin{figure}
\centering
\includegraphics[width=0.70\linewidth]{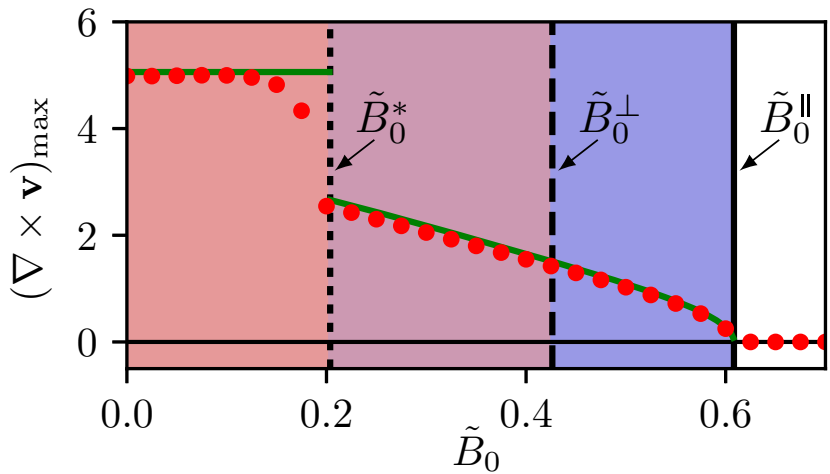}
\caption{Maximum of the vorticity as a function of the external field strength for $\lambda = 0.1$, $a = 0.8$ and $b = 0.1$. The green lines denote the stationary solutions of the amplitude equations~(\ref{eq: amplitude equations vortex lattice}) and (\ref{eq: amplitude equation stripes}), respectively. The red dots are obtained by  numerically solving the full dynamical equation~(\ref{eq: dynamic equation scaled}). Decreasing the external field strength from the onset of the instability at $\tilde{B}^\para_0$ we observe a stripe pattern that becomes unstable with respect to the formation of a square vortex lattice at $\tilde{B}_0^\ast$. Due to the suppression of perpendicular modes by parallel modes, the field strength $\tilde{B}_0^\perp$ obtained by linear stability analysis of the homogeneous stationary solution is not relevant.}
\label{fig: amplitude}
\end{figure}

However, if the strength of the advection term determined by the parameter $\lambda$ in equation~(\ref{eq: dynamic equation scaled}) is small, the full nonlinear nature of this term is expected to be less relevant.
In this case, we expect the formation of a regular square vortex lattice for $\tilde{B}_0 = 0$ and traveling stripes for $\tilde{B}_0^\perp < \tilde{B}_0 < \tilde{B}_0^\para$ [see figure~\ref{fig: patterns}(a) and (b), respectively].
Interestingly, when solving the dynamical equation~(\ref{eq: dynamic equation scaled}) numerically for small values of $\lambda$, we observe a regular square vortex lattice also in the presence of a finite external orienting field, i.e., $\tilde{B}_0 > 0$, provided that the field's magnitude is sufficiently small.
Note that the formation of an asymmetric lattice, that is a lattice where the two directions exhibit different characteristic lengths, is \textit{not} observed in simulations.
This is due to the fact that the external field does not change the critical mode but only shifts the entire growth rate curve (see also section \ref{sec: state diagram}).
Starting from the vortex lattice and increasing the field strength, a transition to regular stripes traveling in the field's direction occurs at a critical strength $\tilde{B}^\ast_0$.
This transition can be observed in figure~\ref{fig: amplitude} where the maximum vorticity [in the rescaled model simply given by the sum of amplitudes of parallel and perpendicular modes, $(\nabla\times\mathbf{v})_\mathrm{max} = 2 A_\para + 2 A_\perp$] is plotted over the external field strength. 
Numerical results, obtained by solving equation~(\ref{eq: dynamic equation scaled}), are denoted by red dots.
The green lines are given by the stationary solutions of the amplitude equations for the square vortex lattice and the traveling stripes, equation~(\ref{eq: amplitude equations vortex lattice}) and (\ref{eq: amplitude equation stripes}), respectively.
Interestingly, the transition field strength $\tilde{B}_0^\ast$ is \textit{not} given by $\tilde{B}^\perp_0$, that is, the field strength where, according to linear stability analysis of the homogeneous stationary solution, perpendicular modes start to grow (see section~\ref{sec: linear stability analisys}).
The reason for this discrepancy is that perpendicular modes become suppressed by parallel modes already at field strengths smaller than $\tilde{B}^\perp_0$.
The transition field strength $\tilde{B}^\ast_0$ is obtained by taking the coupling between the modes into account and performing a linear stability analysis of the stripe pattern with respect to perpendicular modes (see~\ref{app: stability of stripes}).
In order to check for dependencies on the initial values in the numerically obtained data points in figure~\ref{fig: amplitude}, we performed the numerical solution twice, starting from a regular vortex lattice and a stripe pattern.
We found no impact of the chosen initial values indicating the absence of hysteretic behavior.
As visible in figure~\ref{fig: amplitude}, once the vortex lattice is fully developed, the saturated value is given by equations~(\ref{eq: amplitude equations vortex lattice}), which means the external field has no influence on the amplitude.
Instead, preliminary numerical results show that the external field only distorts the lattice.
This intriguing effect will be discussed elsewhere.

\section{Conclusions}
\label{sec: conclusions}

This article studies the impact of a homogeneous, stationary external field on pattern formation in suspensions of microswimmers that exhibit mesoscale turbulence in the field-free case.
Based on a numerical solution of the \textit{full} model presented in section~\ref{sec: theory} and a linear stability analysis, we observe that, in the presence of an orienting field, the patterns become anisotropic and asymmetric.
From a mathematical point of view, the growth rate of modes becomes dependent on the wavevector's direction due to the broken rotational symmetry.
In particular, modes perpendicular to the applied external field are suppressed for intermediate field strengths leading to the dominance of modes parallel to the field.
The resulting structures can be described as stripes that travel along the field's direction.
For even higher field strengths, the instability is completely suppressed and we observe a homogeneous stationary polar state.

In the second part of the paper, we have presented a weakly nonlinear analysis of a reduced model for the effective microswimmer velocity.
The model is significantly easier to handle analytically, yet still exhibits asymmetric pattern formation, which we have analyzed by deriving the amplitude equations~(\ref{eq: amplitude equations vortex lattice}) and (\ref{eq: amplitude equation stripes}).
Upon an increase of the external field, there is a transition form a square vortex lattice to a traveling stripe pattern.
The regularity of these patterns is strongly dependent on the coefficient $\lambda$, that determines the strength of the nonlinear advection term.
When $\lambda$ is increased, defects are generated and the patterns become less regular.
In this case, the amplitude equations,~(\ref{eq: amplitude equations vortex lattice}) and (\ref{eq: amplitude equation stripes}), lose their validity.
This boils down to the problem already discussed in section~\ref{sec: transition stripes lattice}:
The nonlinear energy transfer between scales leads to a shift of the dominating mode contradicting the ansatz leading to the amplitude equations (see also~\ref{app: shift of the dominating mode}).

However, we can still observe one unique feature of the amplitude equation~(\ref{eq: amplitude equation stripes}) for the stripe pattern, even for large values of $\lambda$: the equation is strongly anisotropic as is reflected by the difference of the diffusion coefficients, $D_\para$ and $D_\perp$.
Indeed, the transport of modulations of the pattern in the direction parallel to the external field is approximately one order of magnitude faster than perpendicular to the field [see also equation~(\ref{eq: diffusion coefficients})].
As a consequence, defects in the patterns, generated by the nonlinear advection term, will be elongated by a factor of $D_\para/D_\perp$ in the field's direction.
This is consistent with numerical observations shown in figure~\ref{fig: snapshots}(c) for the full model.

As explained above, the amplitude equation~(\ref{eq: amplitude equation stripes}) does not incorporate the full nonlinear advection term.
This is apparent from the fact that it corresponds to the real-valued Ginzburg-Landau equation, which does not generate defects.
The complex two-dimensional Ginzburg-Landau equation, however, exhibits a variety of chaotic states including one denoted as defect turbulence~\cite{aranson2002world}.
A preliminary comparison of this state with the dynamics of the amplitude modulations of the stripe patterns observed in the present system shows quite similar spatial structures.
The derivation of an amplitude equation valid even for higher values of $\lambda$ presents an interesting future challenge.

As discussed, the external field induces polar order in the system, which leads to net transport in the field's direction.
This net transport can be modified by emerging patterns, as was demonstrated in a recent model for magnetic microswimmers, where band-like structures reflected in the inhomogeneous density of swimmers decrease the net polarization~\cite{koessel2018controlling}.
We also find an influence of emerging patterns on transport properties in our model that assumes a constant density of microswimmers on the coarse-grained level:
The emerging stripe pattern in the polar order parameter field influences the traveling speed [see equation~\ref{eq: traveling speed modified}].
Moreover, preliminary numerical results show that the mean transport in the system is impacted in a quite complex manner.
Further exploring this feedback will be especially relevant for applications, where transport properties are essential.

Anisotropic patterns emerging in microswimmer suspensions subjected to external fields have indeed already been observed in magnetotactic bacteria~\cite{guell1988hydrodynamic,spormann1987unusual}.
However, in these experiments, band-like structures in the swimmer density were found, whereas in our case, we are dealing with a purely orientational effect.
The stripes, visible in the vorticity, correspond to wave-like patterns in the polar order parameter field.

\section*{Acknowledgments}

We thank the Deutsche Forschungsgemeinschaft for financial support through GRK 1558 and SFB 910 (projects B2 and B5), HE 5995/3 and BA 1222/7.

\appendix
\setcounter{section}{0}

\section{Relation to the microscopic model}
\label{app: derivation}

A detailed derivation of the continuum equations in the absence of an external field, equations~(\ref{eq: P (potential form)}) - (\ref{eq: Stokes}), is given in Ref.~\cite{reinken2018derivation}. 
Extending the derivation towards an external field is quite straightforward.
Here, we will give a short summary of the key points.

The overdamped motion of a swimmer $\sigma$ in an ensemble of $\sigma = 1, \dots,S$ identical swimmers is given by the Langevin equations for the position $\mathbf{X}^\sigma$ and the orientation $\mathbf{N}^\sigma$, respectively,
\begin{eqnarray}
\dot{\mathbf{X}}^\sigma &= v_0 \mathbf{N}^\sigma + \mathbf{u}(\mathbf{X}^\sigma) + \sqrt{2 D}\bxi^\sigma,\label{eq: Langevin position}\\
\dot{\mathbf{N}}^\sigma &= \bOmega(\mathbf{X}^\sigma)\cdot\mathbf{N}^\sigma + \bPi \cdot \Big[ a_0\bSigma(\mathbf{X}^\sigma)\cdot\mathbf{N}^\sigma - \nabla_{\mathbf{N}^\sigma}\Phi + \sqrt{1/\tau}\bfeta^\sigma\Big] ,\label{eq: Langevin orientation}
\end{eqnarray}
where the functions $\bxi^\sigma$ and $\bfeta^\sigma$ denote Gaussian white noise generating diffusion in the translational and rotational motion.
Note, that the overdamped Langevin equations are already divided by the translational and rotational friction coefficients, respectively.
Self-propulsion is introduced via the first term on the right-hand side of equation~(\ref{eq: Langevin position}), involving the self-swimming speed $v_0$.
Additionally, the swimmer is transported by advection with the averaged surrounding flow field $\mathbf{u}(\mathbf{x},t)$.
The orientational motion given by equation~(\ref{eq: Langevin orientation}) is determined, first, by rotation and alignment with the flow field.
This is reflected by the terms involving the vorticity $\mathbf{\Omega} = \frac{1}{2}\big[ (\nabla \mathbf{u})^\mathrm{T} - (\nabla \mathbf{u}) \big]$ and deformation rate $\mathbf{\Sigma} = \frac{1}{2}\big[ (\nabla \mathbf{u})^\mathrm{T} + (\nabla \mathbf{u}) \big]$, respectively.
In analogy to Jeffrey's theory for oscillatory tumbling motion of elongated particles in flow~\cite{jeffery1922motion,hinch1979rotation,pedley1992hydrodynamic}, the shape parameter $a_0$ is given as a function of the aspect ratio, defined by length $\ell$ and diameter $d$ of the swimmers,
\begin{equation}
\label{eq: shape parameter}
a_0 = \frac{(\ell/d)^2 - 1}{(\ell/d)^2 +1}.
\end{equation}
Further, the projector $\bPi = \mathbf{I} - \mathbf{N}^\sigma \mathbf{N}^\sigma$ (where $\mathbf{I}$ is the unit matrix) is introduced to conserve the length of the unit vector $\mathbf{N}^\sigma$.
The second deterministic contribution to the orientational motion is the conservative potential $\Phi$ involving all swimmers. 
In the present study we consider both, pair interactions and an external contribution, that is, 
\begin{equation}
\label{eq: potential}
\Phi = \sum_{\mu,\nu} \Phi_\mathrm{int}(\mathbf{N}^\mu,\mathbf{N}^\nu,r_{\mu,\nu}) + \sum_{\mu} \Phi_\mathrm{ext}(\mathbf{N}^\mu).
\end{equation}
The pair potential $\Phi_\mathrm{int}(\mathbf{N}^\mu,\mathbf{N}^\nu,r_{\mu,\nu})$ describes an activity-driven polar alignment of two swimmers with distance $r_{\mu,\nu} = |\mathbf{X}_\mu - \mathbf{X}_\nu| \geq r$,
\begin{equation}
\label{eq: interaction potential}
\Phi_\mathrm{int}(\mathbf{N}^\mu,\mathbf{N}^\nu,r_{\mu,\nu}) = -\frac{\gamma_0 v_0}{2}\mathbf{N}^\mu\cdot\mathbf{N}^\nu \Theta(r - r_{\mu,\nu}),
\end{equation}
where $\gamma_0$ is the strength of the interaction, and $\Theta(x) = 1$ for $x \geq 0$ and zero otherwise.
Note that this is a simplified ansatz that describes the polar alignment of neighboring swimmers due to near-field hydrodynamics interactions~\cite{hoell2018particle}, an effect that has been observed experimentally~\cite{zhang2010collective}.
Finally, the external potential for swimmer $\mu$ is given by
\begin{equation}
\label{eq: external potential}
\Phi_\mathrm{ext}(\mathbf{N}^\mu) = - B_0 \frac{v_0}{\ell}\mathbf{h}\cdot\mathbf{N}^\mu,
\end{equation}
where the unit vector $\mathbf{h}$ denotes the field's direction.
Here, we already introduced the potential in such a way that the field's magnitude relative to the active time scale $\ell/v_0$ defines the dimensionless external field strength $B_0$.
The dimension of the full prefactor $B_0 v_0/\ell$ appearing in equation~(\ref{eq: external potential}) then is an inverse time due to the scaling of the Langevin equations~(\ref{eq: Langevin position}) and (\ref{eq: Langevin orientation}).

Coming back to equations~(\ref{eq: Langevin position}) and (\ref{eq: Langevin orientation}), far-field hydrodynamic interactions are taken into account by the coupling terms involving the average surrounding flow field $\mathbf{u}$.
At low Reynolds numbers, this field is determined by the Stokes equation, augmented by an active contribution to the stress tensor, which, in turn depends on the order parameter.
This coupling eventually leads to equation~(\ref{eq: Stokes}).
Due to the lengthy derivation of the active stress we refer to our previous publication~\cite{reinken2018derivation} for details.

The next step is to obtain the Fokker-Planck equation for the one-particle probability density function $\mathcal{P}(\mathbf{x},\mathbf{n},t)$, which gives the probability to find a swimmer at position $\mathbf{x}$ with orientation $\mathbf{n}$ at time $t$.
To this end, we assume a constant swimmer density $\rho(\mathbf{x},t) = \rho$ and employ a mean-field approximation to treat the two-particle correlations stemming from conservative interactions.
We then project onto orientational moments $\overline{\mathbf{n}}$, $\overline{\mathbf{n}\mathbf{n}}$, ... of $\mathcal{P}(\mathbf{x},\mathbf{n},t)$, which are directly connected to the polar order parameter $\mathbf{P}$ and the nematic order parameter $\mathbf{Q}$ via
\begin{equation}
\label{eq: order parameters}
\mathbf{P} = \overline{\mathbf{n}}, \qquad \mathbf{Q} = \overline{\mathbf{n}\mathbf{n}} - \mathbf{I}/3.
\end{equation}
Following our previous studies~\cite{heidenreich2016hydrodynamic,reinken2018derivation} we apply a closure relation for the nematic order parameter,
\begin{equation}
\label{eq: closure for nematic order parameter}
\mathbf{Q} = q (\mathbf{P} \mathbf{P} - (\mathbf{P} \cdot \mathbf{P})\mathbf{I}/3) + \lambda_\mathrm{K} \bSigma,
\end{equation}
where the coefficients $q$ and $\lambda_\mathrm{K}$ can be calculated analytically~\cite{reinken2018derivation}.
This is an extension of the Doi closure for passive particles, which incorporates the fact that active particles always generate flow gradients affecting the local nematic order.
For moments higher than the second we apply the so-called Hand-closure~\cite{kroeger2008consistent}.
The details of the procedure are discussed in~\cite{reinken2018derivation}.

Finally, we arrive at a closed dynamic equation for the polar order parameter $\mathbf{P}$, given by equation~(\ref{eq: P (potential form) plus field}) that includes a coupling to the Stokes equation~(\ref{eq: Stokes}). The two coefficients $c_\mathrm{I}$ and $c_\mathrm{F}$, not already specified in the main text, are given by
\begin{equation}
c_\mathrm{I} = \frac{8}{9}\tau\rho\gamma_0 v_0 r^3, \qquad c_\mathrm{F} = \frac{f_0 \rho \ell^2}{10 \mu v_0},
\end{equation}
where $f_0$ denotes the strength of the force dipole exerted by every swimmer on the surrounding fluid, and $\mu$ is the effective viscosity of the suspension~\cite{reinken2018derivation}.

\section{Numerical methods}
\label{app: numerical methods}

Our numerical results are obtained by solving the dynamical equations~(\ref{eq: P (potential form) plus field}) or (\ref{eq: dynamic equation scaled}) using the Runge--Kutta--Fehlberg method (RKF45)~\cite{fehlberg1969low}.
We use a finite-difference discretization of the spatial derivatives on a periodic grid consisting of $256\times256$ points.
For the purpose of visualization, we have increased the resolution in figure~\ref{fig: snapshots} by cubic interpolation.
The incompressibility condition is enforced by a pressure-correction method~\cite{pozrikidis2011introduction}.
In the case of the full model, the Stokes equation~(\ref{eq: Stokes}) is solved applying a stream-function approach~\cite{pozrikidis2011introduction}. 
If not otherwise stated, we apply the homogeneous stationary solution introduced in section~\ref{sec: homogeneous stationary solution} as initial conditions and add small random variations.

\section{Details of the linear stability analysis}
\label{app: linear stability analysis}

In the following, we present the calculation of the complex growth rate $\sigma$ [see equations~(\ref{eq: growth rate real}) and (\ref{eq: growth rate imaginary})] determining the stability of the homogeneous stationary solution. 
To this end, we will first eliminate the dependence of the linearized system on the perturbations $\delta \hat{\mathbf{u}}$, $\delta \hat{p}$ and $\delta \hat{q}$, thus reducing the set of variables to just the perturbation of the polar order parameter, $\delta \hat{\mathbf{P}}$.

As a first step we insert the perturbed solution [see equation~(\ref{eq: add perturbations}) and (\ref{eq: perturbations form}) in the main text] into the Stokes equation~(\ref{eq: Stokes}) and linearize, yielding
\begin{equation}
\label{eq: LSA stokes}
-|\mathbf{k}|^2 \delta \hat{\mathbf{u}} = c_\mathrm{F} \bigg(\rmi6 c_\mathrm{I}  k_\para  P_0 \delta \hat{\mathbf{P}} - |\mathbf{k}|^2 \delta \hat{\mathbf{P}} + \frac{1}{28}|\mathbf{k}|^4 \delta \hat{\mathbf{P}}\bigg) + i\mathbf{k} \delta \hat{p}.
\end{equation}
We multiply equation~(\ref{eq: LSA stokes}) by the wavevector $\mathbf{k}$ and employ the incompressibility conditions which, for the perturbed system, yield $\mathbf{k}\cdot \delta\hat{\mathbf{u}} = 0$ and $\mathbf{k}\cdot \delta\hat{\mathbf{P}} = 0$.
From this, we find that the pressure perturbation amplitude must satisfy $\delta \hat{p} = 0$.
Equation~(\ref{eq: LSA stokes}) then yields the perturbation $\delta \hat{\mathbf{u}}$ as a function of the perturbation of the polar order parameter,
\begin{equation}
\label{eq: LSA velocity perturbation}
\delta \hat{\mathbf{u}} = c_\mathrm{F} \Bigg( -\rmi\frac{6 c_\mathrm{I} k_\para P_0}{|\mathbf{k}|^2} + 1 - \frac{1}{28} |\mathbf{k}|^2 \Bigg) \delta \hat{\mathbf{P}}.
\end{equation}
We now consider the equation of motion for the polar order parameter $\mathbf{P}$, equation~(\ref{eq: P (potential form) plus field}).
Performing the linearization yields 
\begin{equation}
\eqalign{
\label{eq: LSA full system}
\sigma \delta \hat{P}_\para = &- \rmi \lambda_0 k_\para P_0 \delta \hat{P}_\para - \alpha \delta \hat{P}_\para - 3 \beta P_0^2 \delta \hat{P}_\para - \Gamma_2 |\mathbf{k}|^2 \delta \hat{P}_\para - \Gamma_4 |\mathbf{k}|^4 \delta \hat{P}_\para\\
&- \frac{4}{5} c_\mathrm{I} B_0 P_0 \delta \hat{P}_\para - \rmi\frac{2}{15} a_0 P_\mathrm{r} B_0 k_\para \delta \hat{u}_\para - \rmi k_\para \delta \hat{q} + \rmi \kappa k_\para \delta \hat{u}_\para P_0,\\
\sigma \delta \hat{P}_\perp = &- \rmi \lambda_0 k_\para P_0 \delta \hat{P}_\perp - \alpha \delta \hat{P}_\perp - \beta P_0^2 \delta \hat{P}_\perp - \Gamma_2 |\mathbf{k}|^2 \delta \hat{P}_\perp - \Gamma_4 |\mathbf{k}|^4 \delta \hat{P}_\perp\\
&- \frac{3}{5} c_\mathrm{I} B_0 P_0 \delta \hat{P}_\perp - \rmi \frac{1}{15} a_0 P_\mathrm{r} B_0 (k_\para \delta \hat{u}_\perp + k_\perp \delta \hat{u}_\para) - \rmi k_\perp \delta \hat{q}\\
&+ \rmi\frac{1}{2} \big(k_\para \delta \hat{u}_\perp - k_\perp \delta \hat{u}_\para\big) P_0 + \rmi \frac{1}{2}\kappa\big(k_\para \delta \hat{u}_\perp + k_\perp\delta \hat{u}_\para\big) P_0.}
\end{equation}
We can eliminate the dependence of equation~(\ref{eq: LSA full system}) on the velocity perturbation $\delta\hat{\mathbf{u}}$ by inserting equation~(\ref{eq: LSA velocity perturbation}).
The reduced system can then be written as
\begin{equation}
\label{eq: LSA matrix form before projector}
\sigma \delta \hat{\mathbf{P}} = - i \mathbf{k} \delta \hat{q} + \mathbf{M}(\mathbf{k}) \cdot \delta \hat{\mathbf{P}},
\end{equation}
where the components of the Jacobian matrix $\mathbf{M}(\mathbf{k})$ are given by
\begin{eqnarray}
\label{eq: LSA linearized system matrix}
M_{\para \para}(\mathbf{k}) = &- \alpha - 3\beta P_0^2 - \Gamma_2 |\mathbf{k}|^2  - \Gamma_4 |\mathbf{k}|^4 - \rmi \lambda_0 k_\para P_0 - \frac{4}{5}B_0 c_\mathrm{I} P_0\\
&- \frac{4}{5}P_\mathrm{r}c_\mathrm{F}c_\mathrm{I}a_0 B_0 \frac{k_\para^2}{|\mathbf{k}|^2}P_0 - \rmi \frac{2}{15} a_0  B_0 c_\mathrm{F} P_\mathrm{r} k_\para \Big(1 - |\mathbf{k}|^2/28\Big) \nonumber\\
&+ 6 \kappa c_\mathrm{I} c_\mathrm{F} \frac{k_\para^2}{|\mathbf{k}|^2} P_0^2 + \rmi \kappa c_\mathrm{F} k_\para \Big(1 - |\mathbf{k}|^2/28\Big) P_0, \nonumber\\
M_{\para \perp}(\mathbf{k}) = &\ 0,\\
M_{\perp \para}(\mathbf{k}) = &\ 0, \\
M_{\perp \perp}(\mathbf{k}) = &- \alpha - \beta P_0^2 - \Gamma_2 |\mathbf{k}|^2  - \Gamma_4 |\mathbf{k}|^4 - \rmi \lambda_0 k_\para P_0 - \frac{3}{5}B_0 c_\mathrm{I} P_0\\
&- \frac{2}{5}P_\mathrm{r}c_\mathrm{F}c_\mathrm{I}a_0 B_0 \frac{k_\para^2 - k_\perp^2}{|\mathbf{k}|^2}P_0 \nonumber \\
&- \rmi \frac{1}{15} a_0 B_0 c_\mathrm{F} P_\mathrm{r}\frac{k_\para^2 - k_\perp^2}{k_\para} \Big(1 - |\mathbf{k}|^2/28\Big) \nonumber \\
&+ \frac{3 \kappa c_\mathrm{I} c_\mathrm{F}}{|\mathbf{k}|^2} \Big[k_\para^2(1+\kappa) + k_\perp^2(1-\kappa)\Big] P_0^2 \nonumber \\
&+ \frac{\rmi \kappa c_\mathrm{F}}{2 k_\para} \Big[k_\para^2(1+\kappa)+ k_\perp^2(1-\kappa)\Big] \Big( 1 - |\mathbf{k}|^2/28\Big) P_0.\nonumber
\end{eqnarray}
Equation~(\ref{eq: LSA matrix form before projector}) still involves the perturbation $\delta \hat{q}$ of the Lagrange multiplier.
Utilizing again the incompressibility condition for the polar order parameter field $\mathbf{P}$, yielding $\mathbf{k}\cdot \delta\hat{\mathbf{P}} = 0$, and equation~(\ref{eq: LSA matrix form before projector}), we find that the perturbation $\delta \hat{q}$ must satisfy
\begin{equation}
\label{eq: LSA Lagrange multiplier perturbation}
\delta \hat{q} = - \rmi\frac{\mathbf{k}}{|\mathbf{k}|^2} \cdot \mathbf{M} \cdot \delta \hat{\mathbf{P}}.
\end{equation}
Inserting equation~(\ref{eq: LSA Lagrange multiplier perturbation}) into equation~(\ref{eq: LSA matrix form before projector}) we can identify the projector $\mathbf{\Pi}(\mathbf{k}) = \mathbf{I} - \mathbf{k}\mathbf{k}/|\mathbf{k}|^2 $. 
With this, we obtain equation~(\ref{eq: linearized system}) in the main text, giving a linearized system involving only perturbations of $\mathbf{P}$.
The complex growth rate can now be readily calculated as solution of the eigenvalue problem [equation~(\ref{eq: linearized system})] via
\begin{equation}
\label{eq: LSA eigenvalues general}
\sigma(\mathbf{k}) = \frac{k_\perp^2}{|\mathbf{k}|^2}M_{\para \para} + \frac{k_\para^2}{|\mathbf{k}|^2}M_{\perp \perp} - \frac{k_\para k_\perp}{|\mathbf{k}|^2}\Big(M_{\para \perp} + M_{\perp \para}\Big).
\end{equation}
For the explicit form of $\sigma(\mathbf{k})$, see equations~(\ref{eq: growth rate real}) and (\ref{eq: growth rate imaginary}) in the main text.

\section{Coefficients in the reduced and rescaled model}
\label{app: coefficients reduced model}

Four dimensionless coefficients remain in the reduced and rescaled model equation~(\ref{eq: dynamic equation scaled}).
These are related to the coefficients of the full model given by equations~(\ref{eq: P (potential form)}), (\ref{eq: functional}), (\ref{eq: convective derivative}) and (\ref{eq: external field}) via
\begin{eqnarray}
\lambda &= \sqrt{\frac{5}{3c_\mathrm{I}}} \sqrt{\frac{-2 \Gamma_4}{\Gamma_2}} \frac{1}{\Gamma_2} \lambda_0,\\
a &= 1 - \frac{4\Gamma_4}{\Gamma_2^2} \alpha,\\
b &= \frac{4\Gamma_4}{\Gamma_2^2} \frac{5}{3c_\mathrm{I}} \beta,\\
\tilde{B}_0 &= \frac{2}{3} \frac{4 \Gamma_4}{\Gamma_2^2} \sqrt{\frac{3 c_\mathrm{I}}{5}} B_0.
\end{eqnarray}

\section{Details of the weakly nonlinear analysis}
\label{app: weakly nonlinear analysis}

Here we present technical details for the weakly nonlinear analysis for the case $\tilde{B}_0 > 0$.
The analysis for the case $\tilde{B}_0 = 0$ is analogous, but less complex due to the overall rotational symmetry.
In fact, the only non-standard feature in the field-free case is the coupling between the two leading-order amplitudes, see equation~(\ref{eq: amplitude equations vortex lattice}).

As introduced in section~\ref{sec: weakly nonlinear analysis field}, the parameter $\varepsilon$ denotes the distance to the bifurcation occurring at $\tilde{B}_0 = \tilde{B}^\para_0$ (when starting from large fields, where the system is homogeneous).
At $\tilde{B}_0 = \tilde{B}^\para_0$, the instability sets in, that is, perturbations with a parallel wavevector start to grow and we observe a transition to a stripe pattern.
Using the definitions of the long time and space scales, $T = \varepsilon^2 t$ and $\mathbf{X} =  \varepsilon (\mathbf{x} - \mathbf{v}_\mathrm{g} t)$, the derivatives in the dynamic equation~(\ref{eq: dynamic equation scaled}) for the effective velocity $\mathbf{v}$ are replaced by
\begin{equation}
\eqalign{
\label{eq: derivatives replacement}
\partial_t \quad &\rightarrow \quad\partial_t - \varepsilon v^\para_\mathrm{g} \partial_{X_\para} - \varepsilon v^\perp_\mathrm{g} \partial_{X_\perp} + \varepsilon^2 \partial_T,\\
\partial_{x_\para} \quad &\rightarrow \quad \partial_{x_\para} + \varepsilon \partial_{X_\para}, \qquad\quad \partial_{x_\perp} \rightarrow \quad \partial_{x_\perp} + \varepsilon \partial_{X_\perp},\\
\nabla^2 \quad &\rightarrow \quad \partial^2_{x_\para} + \partial^2_{x_\perp} + \varepsilon \Big(2\partial_{x_\para}\partial_{X_\para} + 2\partial_{x_\perp}\partial_{X_\perp}\Big) + \varepsilon^2\Big(\partial^2_{X_\para} + \partial^2_{X_\perp}\Big),\\
\nabla^4 \quad &\rightarrow \quad \partial^4_{x_\para} + \partial^4_{x_\perp} + 2 \partial^2_{x_\para}\partial^2_{x_\perp}\\
&\qquad + \varepsilon \Big( 4\partial^3_{x_\para}\partial_{X_\para} + 4\partial^3_{x_\perp}\partial_{X_\perp} + 4\partial^2_{x_\para}\partial_{x_\perp}\partial_{X_\perp} + 4\partial^2_{x_\perp}\partial_{x_\para}\partial_{X_\para}\Big) \nonumber\\
&\qquad + \varepsilon^2 \Big( 6\partial^2_{x_\para}\partial^2_{X_\para} + 6\partial^2_{x_\perp}\partial^2_{X_\perp} + 2\partial^2_{x_\para}\partial^2_{X_\perp} + 2\partial^2_{x_\perp}\partial^2_{X_\para} + \dots \nonumber \\
&\qquad \dots + 8\partial_{x_\para}\partial_{x_\perp}\partial_{X_\para}\partial_{X_\perp} \Big) + \mathcal{O}(\varepsilon^3). \nonumber}
\end{equation}
Note that for the case $\tilde{B}_0 = 0$ the group velocity vanishes, $v^\para_\mathrm{g} = v^\perp_\mathrm{g} = 0$.
This is due to the rotational symmetry and, thus, the absence of net transport in the system.
In the present case, however, the rotational symmetry is broken and, near the onset of the instability at $\tilde{B}_0^\para$, only perturbations with a parallel wavevector grow.
Therefore, we expand the effective velocity field $\mathbf{v} = (v_\para,v_\perp)$ in orders of $\varepsilon$ incorporating only parallel modes characterized by multiples of the critical wavenumber.
The full expansion is given in equation~(\ref{eq: expanison effective velocity full}) in the main text.
Similarly, we expand the Lagrange multiplier $q$ that enforces the incompressibility constraint using only parallel modes,
\begin{equation}
q(\mathbf{x},t,\mathbf{X},T) = q_0 + \sum_{n=1}^{\infty} \sum_{m=0}^{n} \varepsilon^n q_{n,m}(\mathbf{X},T) \rme^{\rmi m (x_\para - c_\para t)} + \mathrm{c.c.} \, .
\label{eq: Lagrange multiplier expansion}
\end{equation}
Finally, the traveling speed $c^\para$ is also expanded in orders of $\varepsilon$,
\begin{equation}
c^\para = c_0^\para + \sum_{n=1}^{\infty} \varepsilon^n c_n^\para,
\label{eq: traveling speed expansion}
\end{equation}
where the zeroth component is proportional to the homogeneous stationary solution $V_0$, i.e., $c_0^\para = \lambda V_0$ [see equation~(\ref{eq: traveling speed rescaled})].
Now, we insert all expansions [equations~(\ref{eq: expanison effective velocity full}), (\ref{eq: traveling speed expansion}) and (\ref{eq: Lagrange multiplier expansion})] into equation~(\ref{eq: dynamic equation scaled}) and replace all derivatives via equation~(\ref{eq: derivatives replacement}).
Sorting the resulting terms according to powers of $\varepsilon$ ($n$) and modes ($m$) we obtain solvability conditions.
In the following, we successively go through the resulting equations:

\begin{itemize}
\item In zeroth order, $\mathcal{O}(\varepsilon^0)$, we recover equation~(\ref{eq: homogeneous stationary solution scaled}) which determines the homogeneous stationary solution $V_0$.

\item In first order, $\mathcal{O}(\varepsilon^1)$, we find that the contributions $v_{1,0}^\para$, $v_{1,1}^\para$, $v_{1,0}^\perp$, $q_{1,0}$, $q_{1,1}$ and $c_1^\para$ must vanish to satisfy simultaneously the dynamical equation~(\ref{eq: dynamic equation scaled}) and the incompressibility condition $\nabla \cdot \mathbf{v} = 0$. For $v_{1,1}^\perp$ we obtain
\begin{equation}
0 = ( a - b V_0^2 - \frac{3}{2}\tilde{B}_0 V_0) v_{1,1}^\perp = \varepsilon^2 v_{1,1}^\perp,
\label{eq: WNLA first order}
\end{equation}
where we used the definition of $\varepsilon$ according to equation~(\ref{eq: bifurcation point}).
Thus, we find that $v_{1,1}^\perp$ actually contributes in $\mathcal{O}(\varepsilon^3)$ to the amplitude equation obtained below.

\item In second order, $\mathcal{O}(\varepsilon^2)$, using the incompressibility constraint, we find the relation
\begin{equation}
v_{2,1}^\para = \rmi\partial_{X_\perp}v_{1,1}^\perp.
\label{eq: WNLA second order incompressibility}
\end{equation}
As a solvability condition for equation~(\ref{eq: dynamic equation scaled}) we obtain for the zeroth mode ($m=0$)
\begin{equation}
v_{2,0}^\para = D_0 |v_{1,1}^\perp|^2,
\label{eq: WNLA second order zeroth mode}
\end{equation}
where 
\begin{equation}
D_0 = \frac{- 2 b V_0 + \tilde{B}_0}{1 + 2 b V_0^2 + \frac{1}{2} \tilde{B}_0 V_0}.
\label{eq: relation for D_0}
\end{equation}
Matching all terms containing the first mode ($m=1$) yields
\begin{equation}
q_{2,1} = \rmi D_\perp v_{2,1}^\para = - D_\perp \partial_{X_\perp}v_{1,1}^\perp,
\label{eq: WNLA second order first mode}
\end{equation}
where we inserted equation~(\ref{eq: WNLA second order incompressibility}).
The coefficient $D_\perp$ is given in equation~(\ref{eq: diffusion coefficients}) in the main text.
Equations~(\ref{eq: WNLA second order incompressibility}) - (\ref{eq: WNLA second order first mode}) show that the second order amplitudes $v_{2,0}^\para$, $v_{2,1}^\para$ and $q_{2,1}$ are ``slaved'' to the first order amplitude $v_{1,1}^\perp$.
Further, we obtain the second order contribution to the traveling speed
\begin{equation}
\label{eq: traveling speed second order}
c_2^\para = \lambda D_0 |v_{1,1}^\perp|^2.
\end{equation}
All other higher order modes ($n>1$), i.e., $v_{2,2}^\para$, $v_{2,0}^\perp$, $v_{2,1}^\perp$, $v_{2,2}^\perp$, $q_{2,0}$, $q_{2,2}$, $\dots$ either violate the incompressibility constraint or the solvability conditions obtained from equation~(\ref{eq: dynamic equation scaled}), or they contribute in orders higher than $\mathcal{O}(\varepsilon^3)$.

\item In third order, $\mathcal{O}(\varepsilon^3)$, we finally obtain a dynamic equation for the amplitude $v_{1,1}^\perp$ by matching terms containing the first mode ($m=1$),
\begin{equation}
\eqalign{
\label{eq: amplitude equation stripes preliminary step 1}
\partial_T v_{1,1}^\perp  = & + v_{1,1}^\perp - 3 b |v_{1,1}^\perp|^2 v_{1,1}^\perp - 2 b V_0 v_{2,0}^\para v_{1,1}^\perp - \frac{3}{2}B_0 v_{2,0}^\para v_{1,1}^\perp\\
&+ 4 \partial_{X_\para}^2 v_{1,1}^\perp - \partial_{X_\perp} q_{2,1}^\para.}
\end{equation}
Inserting equations~(\ref{eq: WNLA second order zeroth mode}) and (\ref{eq: WNLA second order first mode}) into equation~(\ref{eq: amplitude equation stripes preliminary step 1}) and replacing $v_{1,1}^\perp \rightarrow A$ yields
\begin{equation}
\label{eq: amplitude equation stripes preliminary step 2}
\partial_T A = A - g |A|^2 A + D_\para \partial_{X_\para}^2 A + D_\perp \partial_{X_\perp}^2 A.
\end{equation}
The diffusion coefficients $D_\para$ and $D_\perp$ are given in equation~(\ref{eq: diffusion coefficients}) in the main text.
Further, the coefficient of the cubic term follows as
\begin{equation}
\label{eq: cubig coefficient g}
g = 3 b+ 2 b V_0 D_0 + \frac{3}{2} \tilde{B}_0 D_0.
\end{equation}
\end{itemize}
Finally, scaling back to the fast time and length scales, $t$ and $\mathbf{x}$, yields the amplitude equation~(\ref{eq: amplitude equation stripes}) in the main text.

\section{Shift of the dominating mode}
\label{app: shift of the dominating mode}

\begin{figure}
\centering
\includegraphics[width=0.65\linewidth]{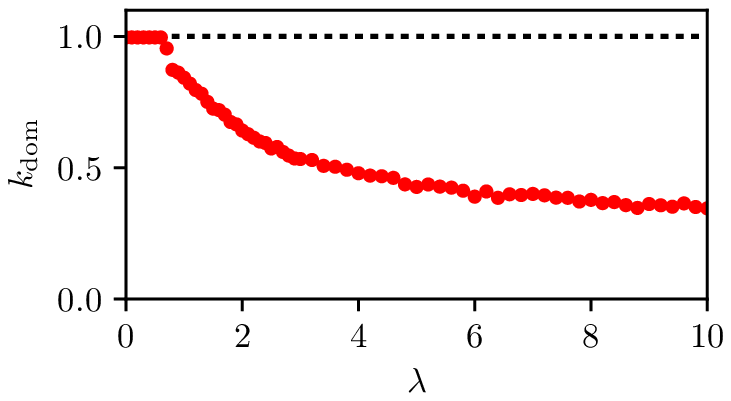}
\caption{Dominating mode $k_\mathrm{dom}$ in the reduced system [equation~(\ref{eq: dynamic equation scaled})] as function of the strength $\lambda$ of the advection term for the field-free case, $\tilde{B}_0 = 0$. The remaining parameters are $a = 0.8$ and $b = 0.1$. The data points are obtained by numerically solving [equation~(\ref{eq: dynamic equation scaled})] and determining the first maximum $\Lambda_\mathrm{max}$ of the spatial correlation function $\big\langle \mathbf{v}(\mathbf{x}) \cdot \mathbf{v}(\mathbf{x} + \Delta\mathbf{x})\big\rangle$ (where $\big\langle \dots \big\rangle$ denotes the spatial and temporal average). The dominant mode then follows as $k_\mathrm{dom} = 2 \pi/\Lambda_\mathrm{max}$. To reduce the number of transient defects remaining from the initial conditions, we start from a regular vortex lattice with small random variations. For small $\lambda$, the characteristic length scale of the patterns is given by the critical mode $k_\mathrm{c} = 1$. For larger $\lambda$, turbulent motion sets in and energy is transfered to larger scales, i.e., smaller wavenumbers.}
\label{fig: mode shift}
\end{figure}

As discussed in sections~\ref{sec: transition stripes lattice}, the nonlinear advection term leads to energy transfer between different scales once its strength, $\lambda$, is large enough. 
As a consequence, the dominating mode in the system shifts. This shift is towards larger scales, i.e., smaller wavenumbers, for a two-dimensional system~\cite{davidson2015turbulence,james2018turbulence}. 
To illustrate this point, we plot the dominating mode as function of the strength of the advection term $\lambda$ in figure~\ref{fig: mode shift}.
The data points are obtained numerically on the basis of the reduced and rescaled model equation~(\ref{eq: dynamic equation scaled}).

\section{Stability of stripe pattern}
\label{app: stability of stripes}

The emerging pattern near the bifurcation at $\tilde{B}_0^\para$ manifests itself as stripes in the vorticity.
With decreasing external field strength, the stripe pattern becomes unstable against the formation of a square vortex lattice.
The corresponding critical field strength can be obtained by performing a suitable linear stability analysis.
In contrast to section \ref{sec: linear stability analisys}, we here add a perturbation to the stripe pattern characterized by the stationary solution $A = \sqrt{\sigma_\mathrm{Re}^\para / g}$ of the amplitude equation~(\ref{eq: amplitude equation stripes}) (and not to the homogeneous stationary solution $V_0$).
The full stripe pattern solution on the level of the collective microswimmer velocity  $\mathbf{v}$ is given as
\begin{equation}
\label{eq: stripe pattern}
\eqalign{
v_\para = V_0 + D_0 |A|^2 ,\\
v_\perp = A \rme^{\rmi (x_\para - c_\para t)} + A^\ast \rme^{-\rmi (x_\para - c_\para t)},}
\end{equation}
where $A^\ast$ denotes the complex conjugate of $A$.
The form of the perturbation we consider is a perpendicular mode with critical wavenumber $k_\mathrm{c} = 1$, i.e.,
\begin{equation}
\label{eq: stripes plus perturbation}
\eqalign{
v_\para = V_0 + D_0 |A|^2 + \delta\hat{v}_\para \rme^{\tilde{\sigma}_\perp t + \rmi x_\perp},\\
v_\perp = A \rme^{\rmi (x_\para - c_\para t)} + A^\ast \rme^{-\rmi (x_\para - c_\para t)} + \delta\hat{v}_\perp \rme^{\tilde{\sigma}_\perp t + \rmi x_\perp}.}
\end{equation}
Utilizing the incompressibility condition $\nabla \cdot \mathbf{v} = 0$, we find $\delta\hat{v}_\perp = 0$.
We insert equation~(\ref{eq: stripes plus perturbation}) into the dynamic equation for $\mathbf{v}$ [equation~(\ref{eq: dynamic equation scaled})] and obtain, after linearization, the growth rate of perpendicular modes with critical wavenumber $k_\mathrm{c} = 1$,
\begin{equation}
\label{eq: growth rate perpendicular mode}
\eqalign{
\tilde{\sigma}_\perp = a &- b(3 V_0^2 + 3 D_0^2 |A|^4 + 6 P_0 D_0 |A|^2 + 2 |A^2|) \\ 
&- 2 B_0(V_0 + D_0 |A|^2).}
\end{equation}
The critical field strength $\tilde{B}^\ast_0$ defining the transition from the stripe pattern to the vortex lattice is then obtained by setting $\tilde{\sigma}_\perp = 0$.
Note that $\tilde{\sigma}_\perp$ depends on the amplitude of the parallel modes, $A$.
This coupling is the reason why the field strength at the transition, $\tilde{B}^\ast_0$, is smaller than $\tilde{B}_0^\perp$.



\begin{thebibliography}{10}
\expandafter\ifx\csname url\endcsname\relax
  \def\url#1{{\tt #1}}\fi
\expandafter\ifx\csname urlprefix\endcsname\relax\def\urlprefix{URL }\fi
\providecommand{\eprint}[2][]{\url{#2}}

\bibitem{buttinoni2013dynamical}
Buttinoni I, Bialk{\'e} J, K{\"u}mmel F, L{\"o}wen H, Bechinger C and Speck T
  2013 {\em Phys. Rev. Lett.\/} {\bf 110} 238301

\bibitem{peruani2012collective}
Peruani F, Starru{\ss} J, Jakovljevic V, S{\o}gaard-Andersen L, Deutsch A and
  B{\"a}r M 2012 {\em Phys. Rev. Lett.\/} {\bf 108} 098102

\bibitem{zhang2010collective}
Zhang H~P, Be'er A, Florin E~L and Swinney H~L 2010 {\em Proc. Natl. Acad. Sci.
  U.S.A.\/} {\bf 107} 13626--13630

\bibitem{schaller2010polar}
Schaller V, Weber C, Semmrich C, Frey E and Bausch A~R 2010 {\em Nature\/} {\bf
  467} 73--77

\bibitem{sumino2012large}
Sumino Y, Nagai K~H, Shitaka Y, Tanaka D, Yoshikawa K, Chat{\'e} H and Oiwa K
  2012 {\em Nature\/} {\bf 483} 448--452

\bibitem{rabani2013collective}
Rabani A, Ariel G and Be'er A 2013 {\em PloS One\/} {\bf 8} e83760

\bibitem{vicsek1995novel}
Vicsek T, Czir{\'o}k A, Ben-Jacob E, Cohen I and Shochet O 1995 {\em Phys. Rev.
  Lett.\/} {\bf 75} 1226

\bibitem{stenhammer2017role}
Stenhammar J, Nardini C, Nash R~W, Marenduzzo D and Morozov A 2017 {\em Phys.
  Rev. Lett.\/} {\bf 119}(2) 028005

\bibitem{simha2002hydrodynamic}
Simha R~A and Ramaswamy S 2002 {\em Phys. Rev. Lett.\/} {\bf 89} 058101

\bibitem{ramaswamy2003active}
Ramaswamy S, Simha R~A and Toner J 2003 {\em Europhys. Lett.\/} {\bf 62} 196

\bibitem{toner2005hydrodynamics}
Toner J, Tu Y and Ramaswamy S 2005 {\em Ann. Phys.\/} {\bf 318} 170--244

\bibitem{cates2008shearing}
Cates M, Fielding S, Marenduzzo D, Orlandini E and Yeomans J 2008 {\em Phys.
  Rev. Lett.\/} {\bf 101} 068102

\bibitem{heidenreich2011nonlinear}
Heidenreich S, Hess S and Klapp S~H 2011 {\em Phys. Rev. E\/} {\bf 83} 011907

\bibitem{giomi2015geometry}
Giomi L 2015 {\em Phys. Rev. X\/} {\bf 5} 031003

\bibitem{saintillan2008instabilities}
Saintillan D and Shelley M~J 2008 {\em Phys. Rev. Lett.\/} {\bf 100} 178103

\bibitem{saintillan2009active}
Saintillan D and Shelley M~J 2013 {\em C. R. Physique\/} {\bf 14} 497--517

\bibitem{baskaran2008hydrodynamics}
Baskaran A and Marchetti M~C 2008 {\em Phys. Rev. E\/} {\bf 77} 011920

\bibitem{heidenreich2016hydrodynamic}
Heidenreich S, Dunkel J, Klapp S~H~L and B{\"a}r M 2016 {\em Phys. Rev. E\/}
  {\bf 94} 020601

\bibitem{reinken2018derivation}
Reinken H, Klapp S~H, B{\"a}r M and Heidenreich S 2018 {\em Phys. Rev. E\/}
  {\bf 97} 022613

\bibitem{lauga2009hydrodynamics}
Lauga E and Powers T~R 2009 {\em Rep. Prog. Phys.\/} {\bf 72} 096601

\bibitem{ramaswamy2010mechanics}
Ramaswamy S 2010 {\em Annu. Rev. Condens. Matter Phys.\/} {\bf 1} 323--345

\bibitem{romanczuk2012active}
Romanczuk P, B{\"a}r M, Ebeling W, Lindner B and Schimansky-Geier L 2012 {\em
  Eur. Phys. J. Spec. Top.\/} {\bf 202} 1--162

\bibitem{Marchetti2013}
Marchetti M, Joanny J, Ramaswamy S, Liverpool T, Prost J, Rao M and Simha R~A
  2013 {\em Rev. Mod. Phys.\/} {\bf 85} 1143

\bibitem{Elgeti2015}
Elgeti J, Winkler R~G and Gompper G 2015 {\em Rep. Prog. Phys.\/} {\bf 78}
  056601

\bibitem{menzel2015tuned}
Menzel A~M 2015 {\em Phys. Rep.\/} {\bf 554} 1 -- 45

\bibitem{zottl2016emergent}
Z{\"o}ttl A and Stark H 2016 {\em J. Phys. Condens. Matter\/} {\bf 28} 253001

\bibitem{Bechinger2016}
Bechinger C, Di~Leonardo R, L\"owen H, Reichhardt C, Volpe G and Volpe G 2016
  {\em Rev. Mod. Phys.\/} {\bf 88}(4) 045006

\bibitem{klapp2016collective}
Klapp S~H 2016 {\em Curr. Opin. Colloid Interface Sci.\/} {\bf 21} 76--85

\bibitem{eisenbach2004chemotaxis}
Eisenbach M 2004 {\em Chemotaxis\/} (World Scientific Publishing Company)

\bibitem{taktikos2011modeling}
Taktikos J, Zaburdaev V and Stark H 2011 {\em Phys. Rev. E\/} {\bf 84} 041924

\bibitem{garcia2013light}
Garcia X, Rafa{\"\i} S and Peyla P 2013 {\em Phys. Rev. Lett.\/} {\bf 110}
  138106

\bibitem{martin2016photofocusing}
Martin M, Barzyk A, Bertin E, Peyla P and Rafa\"i S 2016 {\em Phys. Rev. E\/}
  {\bf 93} 051101

\bibitem{bazylinski2004magnetosome}
Bazylinski D~A and Frankel R~B 2004 {\em Nat. Rev. Microbiol.\/} {\bf 2} 217

\bibitem{waisbord2016environment}
Waisbord N, Lef{\`e}vre C, Bocquet L, Ybert C and Cottin-Bizonne C 2016 {\em
  arXiv preprint arXiv:1603.00490\/}

\bibitem{nadkarni2013comparison}
Nadkarni R, Barkley S and Fradin C 2013 {\em PLoS One\/} {\bf 8} e82064

\bibitem{popp2014polarity}
Popp F, Armitage J~P and Sch{\"u}ler D 2014 {\em Nat. Commun.\/} {\bf 5} 5398

\bibitem{fukui1985negative}
Fukui K and Asai H 1985 {\em Biophys. J.\/} {\bf 47} 479--482

\bibitem{tenhagen2014gravitaxis}
Ten~Hagen B, K{\"u}mmel F, Wittkowski R, Takagi D, L{\"o}wen H and Bechinger C
  2014 {\em Nat. Commun.\/} {\bf 5} 4829

\bibitem{trivedi2015bacterial}
Trivedi R~R, Maeda R, Abbott N~L, Spagnolie S~E and Weibel D~B 2015 {\em Soft
  matter\/} {\bf 11} 8404--8408

\bibitem{sokolov2015individual}
Sokolov A, Zhou S, Lavrentovich O~D and Aranson I~S 2015 {\em Phys. Rev. E\/}
  {\bf 91} 013009

\bibitem{schwarz2017hybrid}
Schwarz L, Medina-S{\'a}nchez M and Schmidt O~G 2017 {\em Appl. Phys. Rev.\/}
  {\bf 4} 031301

\bibitem{martel2009flagellated}
Martel S, Mohammadi M, Felfoul O, Lu Z and Pouponneau P 2009 {\em Int. J.
  Robotics Res.\/} {\bf 28} 571--582

\bibitem{felfoul2016magneto}
Felfoul O, Mohammadi M, Taherkhani S, De~Lanauze D, Xu Y~Z, Loghin D, Essa S,
  Jancik S, Houle D, Lafleur M {\em et~al.\/} 2016 {\em Nat. Nanotechnol.\/}
  {\bf 11} 941

\bibitem{kaiser2014transport}
Kaiser A, Peshkov A, Sokolov A, ten Hagen B, L{\"o}wen H and Aranson I~S 2014
  {\em Phys. Rev. Lett.\/} {\bf 112} 158101

\bibitem{sokolov2010swimming}
Sokolov A, Apodaca M~M, Grzybowski B~A and Aranson I~S 2010 {\em Proc. Natl.
  Acad. Sci. U.S.A.\/} {\bf 107} 969--974

\bibitem{jalali2015microswimmer}
Jalali M~A, Khoshnood A and Alam M~R 2015 {\em J. Fluid Mech.\/} {\bf 779}
  669--683

\bibitem{wensink2012meso}
Wensink H~H, Dunkel J, Heidenreich S, Drescher K, Goldstein R~E, L{\"o}wen H
  and Yeomans J~M 2012 {\em Proc. Natl. Acad. Sci. U.S.A.\/} {\bf 109}
  14308--14313

\bibitem{dombrowski2004self}
Dombrowski C, Cisneros L, Chatkaew S, Goldstein R~E and Kessler J~O 2004 {\em
  Phys. Rev. Lett.\/} {\bf 93} 098103

\bibitem{zhang2009swarming}
Zhang H, Be'er A, Smith R~S, Florin E~L and Swinney H~L 2009 {\em Europhys.
  Lett.\/} {\bf 87} 48011

\bibitem{aranson2012physical}
Sokolov A and Aranson I~S 2012 {\em Phys. Rev. Lett.\/} {\bf 109} 248109

\bibitem{lushi14fluid}
Lushi E, Wioland H and Goldstein R~E 2014 {\em Proc. Natl. Acad. Sci. U.S.A.\/}
  {\bf 111} 9733--9738

\bibitem{nishiguchi2015mesoscopic}
Nishiguchi D and Sano M 2015 {\em Phys. Rev. E\/} {\bf 92} 052309

\bibitem{davidson2015turbulence}
Davidson P 2015 {\em Turbulence: an introduction for scientists and
  engineers\/} (Oxford University Press)

\bibitem{ilkanaiv2017effect}
Ilkanaiv B, Kearns D~B, Ariel G and Be’er A 2017 {\em Phys. Rev. Lett.\/}
  {\bf 118} 158002

\bibitem{dunkel2013minimal}
Dunkel J, Heidenreich S, B{\"a}r M and Goldstein R~E 2013 {\em New J. Phys.\/}
  {\bf 15} 045016

\bibitem{dunkel2013fluid}
Dunkel J, Heidenreich S, Drescher K, Wensink H~H, B{\"a}r M and Goldstein R~E
  2013 {\em Phys. Rev. Lett.\/} {\bf 110} 228102

\bibitem{slomka2015generalized}
S{\l}omka J and Dunkel J 2015 {\em Eur. Phys. J. ST\/} {\bf 224} 1349--1358

\bibitem{oza2016generalized}
Oza A~U, Heidenreich S and Dunkel J 2016 {\em Eur. Phys. J. E\/} {\bf 39} 97

\bibitem{bratanov2015new}
Bratanov V, Jenko F and Frey E 2015 {\em Proc. Natl. Acad. Sci. U.S.A.\/} {\bf
  112} 15048--15053

\bibitem{james2018vortex}
James M and Wilczek M 2018 {\em Eur. Phys. J. E\/} {\bf 41} 21

\bibitem{james2018turbulence}
James M, Bos W~J and Wilczek M 2018 {\em Phys. Rev. Fluids\/} {\bf 3} 061101(R)

\bibitem{grossmann2014vortex}
Gro{\ss}mann R, Romanczuk P, B{\"a}r M and Schimansky-Geier L 2014 {\em Phys.
  Rev. Lett.\/} {\bf 113} 258104

\bibitem{grossmann2015pattern}
Gro{\ss}mann R, Romanczuk P, B{\"a}r M and Schimansky-Geier L 2015 {\em Eur.
  Phys. J. Special Topics\/} {\bf 224} 1325--1347

\bibitem{toner1998flocks}
Toner J and Tu Y 1998 {\em Phys. Rev. E\/} {\bf 58} 4828

\bibitem{hoell2018particle}
Hoell C, L{\"o}wen H and Menzel A~M 2018 {\em arXiv preprint
  arXiv:1807.08564\/}

\bibitem{blums1997magnetic}
Blums E, Cebers A and Maiorov M~M 1997 {\em Magnetic fluids\/} (Walter de
  Gruyter)

\bibitem{rosensweig2013ferrohydrodynamics}
Rosensweig R~E 2013 {\em Ferrohydrodynamics\/} (Courier Corporation)

\bibitem{koessel2018controlling}
Koessel F~R and Jabbari-Farouji S 2018 {\em arXiv preprint arXiv:1802.07364\/}

\bibitem{edwards2009spontaneous}
Edwards S and Yeomans J 2009 {\em Europhys. Lett.\/} {\bf 85} 18008

\bibitem{cross1993pattern}
Cross M~C and Hohenberg P~C 1993 {\em Rev. Mod. Phys.\/} {\bf 65} 851

\bibitem{newell1993order}
Newell A~C, Passot T and Lega J 1993 {\em Ann. Rev. Fluid Mech.\/} {\bf 25}
  399--453

\bibitem{aranson2002world}
Aranson I~S and Kramer L 2002 {\em Rev. Mod. Phys.\/} {\bf 74} 99

\bibitem{guell1988hydrodynamic}
Guell D, Brenners H, Frankel R~B and Hartman H 1988 {\em Physics\/}  139

\bibitem{spormann1987unusual}
Spormann A~M 1987 {\em FEMS Microbiol. Ecol.\/} {\bf 3} 37--45

\bibitem{jeffery1922motion}
Jeffery G~B 1922 {\em Proc. R. Soc. Lond. A\/} {\bf 102} 161--179

\bibitem{hinch1979rotation}
Hinch E~J and Leal L~G 1979 {\em J. Fluid Mech.\/} {\bf 92} 591 -- 508

\bibitem{pedley1992hydrodynamic}
Pedley T and Kessler J~O 1992 {\em Annu. Rev. Fluid Mech.\/} {\bf 24} 313--358

\bibitem{kroeger2008consistent}
Kr{\"o}ger M, Ammar A and Chinesta F 2008 {\em J. Non-Newtonian Fluid Mech.\/}
  {\bf 149} 50--55

\bibitem{fehlberg1969low}
Fehlberg E 1969 {\em NASA Technical Report\/} {\bf 315}

\bibitem{pozrikidis2011introduction}
Pozrikidis C 2011 {\em Introduction to theoretical and computational fluid
  dynamics\/} (Oxford University Press)

\end{thebibliography}

\providecommand{\newblock}{}

\end{document}